\documentclass[12pt]{article}
\usepackage{amsmath,amssymb}
\usepackage[dvips]{graphicx}

\begin{document}

\begin{titlepage}

  \begin{normalsize}
  \begin{flushright}
    YITP-03-7\\
    hep-th/0302190\\
    February 2003
  \end{flushright}
  \end{normalsize}
  
  \vspace{2cm}
  
  \begin{Large}
  \begin{center}
  \bf
    Notes on the Construction of the D2-brane \\ from Multiple D0-branes
  \end{center}
  \end{Large}
  \vspace{1cm}

\begin{center}
   Yoshifumi Hyakutake
   \footnote{E-mail :hyaku@yukawa.kyoto-u.ac.jp}
   \\
   \vspace{6mm}
   {\it Yukawa Institute for Theoretical Physics, Kyoto University} 
   \\
   {\it Sakyo-ku, Kyoto 606-8502, Japan}
\end{center}

\vspace{2cm}
\begin{center}
  \large{Abstract}
\end{center}

\begin{quote}
\begin{normalsize}
We investigate the correspondence between the
D2-brane, which is described by the abelian Born-Infeld action, 
and multiple D0-branes,
which are done by the nonabelian Born-Infeld action.
We construct effective actions for the fuzzy cylinder,
sphere and plane formed via D0-branes
and compare these actions with those for the cylindrical, 
spherical and planar D2-brane. We show that in the continuous limit,
the effective actions for the fuzzy D0-branes 
precisely coincide with those for the single D2-brane 
if both the fuzziness of the D0-brane and the area occupied per a
unit of magnetic flux on the D2-brane are equal to $(2\pi\ell_s)^2$.
\end{normalsize}
\end{quote}

\end{titlepage}

\setlength{\baselineskip}{0.65cm}

\tableofcontents

\vspace{1cm}
\section{Introduction} \label{sec:Intro}

It is well-known that multiple D0-branes form a fuzzy sphere
in the background of some constant Ramond-Ramond 4-form field 
strength\cite{Mye}. This is what is called the Myers effect.
One of the interesting features of this effect is that the fuzzy sphere
carries the dielectric charge for the R-R 3-form potential.
This means that the fuzzy sphere is actually a bound state
of a D2-brane and D0-branes, and implies that the D2-brane can 
be constructed from multiple D0-branes.

Let us observe the Myers effect in detail.
We consider the case where the fuzzy sphere is formed via $M$ D0-branes.
By evaluating the nonabelian Born-Infeld action and
Chern-Simons action for $M$ D0-branes, we obtain the potential energy
for the fuzzy sphere like
\begin{alignat}{3}
  V_{\text{D}0} &= \sqrt{(MT_0)^2 + (4\pi r^2 T_2)^2(1-\tfrac{1}{M^2})} 
  - \frac{16\pi hT_2 r^3}{3} (1 - \tfrac{1}{M^2}), \label{eq:D0pot}
\end{alignat}
where $-4h(=G^{(4)}_{0123})$ is the constant R-R 4-form field strength and the
parameter $r$ represents a radius of the fuzzy sphere.
$T_0$ is the mass of the D0-brane and $T_2$ is the tension of the D2-brane.
If we take the limit that $M$ is large and $h$ is sufficiently small, 
we can expand the Born-Infeld part of the above potential energy
and find a local minimum near $r = h\lambda M$, where $\lambda$ is defined as
$\lambda = 2\pi \ell_s^2$ and $\ell_s$ is the string length.
Note that the Chern-Simons part of the potential energy
can be approximated as $\tfrac{4\pi(h\lambda M)^3}{3} T_2 G^{(4)}_{0123}$. 
This means that the fuzzy sphere carries the dielectric charge 
for the R-R 3-form potential.

On the other hand, the same dielectric phenomenon can also be confirmed
by analyzing the abelian Born-Infeld action and Chern-Simons action
for a spherical D2-brane.
The potential energy for the spherical D2-brane with
$M$ units of magnetic flux on its world-volume is evaluated as
\begin{alignat}{3}
  V_{\text{D}2} &= \sqrt{(MT_0)^2 + (4\pi r^2 T_2)^2} 
  - \frac{16\pi hT_2 r^3}{3}. \label{eq:D2pot}
\end{alignat}
It is easy to see that the potential energy (\ref{eq:D0pot}) approaches 
to the energy (\ref{eq:D2pot}) as $M$ goes sufficiently large.
Thus the dielectric phenomenon can be viewed from two different ways, 
that is, as the classical configurations of multiple D0-branes which form
the fuzzy sphere or as the classical configuration of the spherical D2-brane 
with magnetic flux on its world-volume.

So far we have observed that the system of the spherical D2-brane
with magnetic flux on the world-volume is realized as the classical 
fuzzy configurations of D0-branes. Now it is natural to pose a question 
whether an effective action for the spherical D2-brane with magnetic flux is 
constructed from the multiple D0-branes which form the fuzzy sphere 
by adding some fluctuations around it.
Especially it is easy to show that the world-volume 
theory on the spherical D2-brane is described by $U(1)$ gauge theory.
So the fluctuations around the fuzzy sphere must also be described by 
$U(1)$ gauge theory if the fuzzy sphere precisely corresponds to the 
spherical D2-brane with magnetic flux in the continuous, or large $M$, limit.

In this paper we analyze the nonabelian Born-Infeld action for D0-branes 
around the backgrounds of the fuzzy cylinder, sphere and 
plane\footnote{For the plane case, see also refs.\,\cite{BFSS} and \cite{VK}.
The construction of D-branes via tachyon condensation is investigated in
ref.\,\cite{AST}.}.
And we show that fluctuations around these fuzzy backgrounds 
are certainly described by $U(1)$ gauge theories. Furthermore we see that 
these effective actions precisely coincide with the actions obtained by 
evaluating the abelian Born-Infeld action for the D2-brane if both
the fuzziness of the D0-brane and the area occupied per a unit of magnetic 
flux on the D2-brane are equal to $(2\pi\ell_s)^2$.

It should be mentioned that in this paper we consider the cylindrical
and spherical D2-brane, or the fuzzy cylinder and sphere
formed via D0-branes, in the background of the flat space-time.
These configurations are unstable against the collapse because of the 
D2-brane tension, however,
in any case it is possible to stabilize those configurations 
by introducing some nontrivial background R-R flux, as in the case
of the Myers effect\footnote{For the stability of D-branes, see also
refs. \cite{NO}-\cite{SS}.}. Since the effects of the background R-R flux appears
through the Chern-Simons action, for the analyses of the Born-Infeld action,
it does not matter whether the background R-R flux exists or not.

The contents of this paper are as follows.
In section 2, we give some comments on properties of the fuzzy cylinder,
sphere and plane by explicitly representing matrices.
In section 3, we obtain the effective actions for the cylindrical, spherical
and planar D2-brane by evaluating the abelian Born-Infeld action.
In section 4, we construct the effective actions for the fuzzy cylinder,
sphere and plane, and compare these actions with the results obtained 
in the section 3. Some discussions are given in section 5.

\vspace{1cm}
\section{Some Comments on Fuzzy Surfaces with Axial Symmetry} \label{sec:Comm}

\vspace{0.3cm}
In this section we review various fuzzy surfaces which are
embedded in the three dimensional flat space $(x^1,x^2,x^3)$.
First we consider a fuzzy cylinder extending to the $x^3$ direction 
and generalize this to a fuzzy surface with axial symmetry around
the $x^3$ direction. Then we discuss a fuzzy sphere and plane in order.

The fuzzy surface is represented by a set of 
three hermitian matrices ($X^1$, $X^2$, $X^3$). 
If these three matrices are simultaneously diagonalized, 
each set of diagonal elements $(X^1_{mm},X^2_{mm},X^3_{mm})$ 
is interpreted as a position in the space $(x^1,x^2,x^3)$. 
Note that these matrices describe just a set of `points' 
and do not represent a `surface'. In order to construct 
a `surface' from matrices, the commutation relations among 
$X^1$, $X^2$ and $X^3$ must be nontrivial, as we will see later.

It is useful to remark that the size of matrices is finite for 
a closed fuzzy surface and infinite for an open one. 
This is because the size of matrices is related to the area of 
the fuzzy surface. The closed fuzzy surface can be approximated as a smooth one
if the size of matrices is large enough. 
Similarly the open fuzzy surface can be approximated 
as a smooth one if the size of partial matrices
assigned to some finite area is sufficiently large.

\vspace{0.3cm}
\subsection{The fuzzy cylinder}

\begin{figure}[tb]
\begin{center}
  \includegraphics[width=9cm,height=4cm,keepaspectratio]{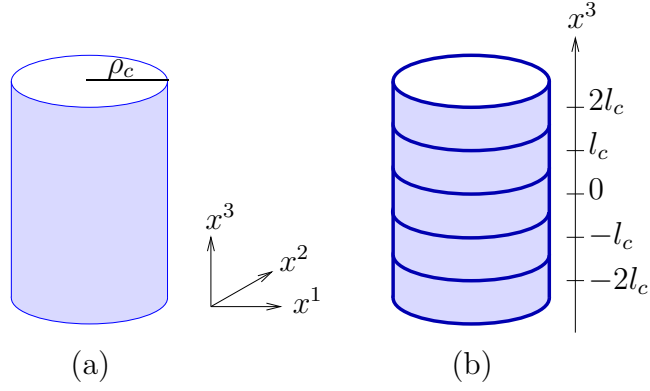}
\begin{picture}(300,0)
  \put(66,0){(a)}
  \put(210,0){(b)}
  \put(150,23){$x^1$}
  \put(145,39){$x^2$}
  \put(117,55){$x^3$}
  \put(254,131){$x^3$}
  \put(262,32){$-2l_c$}
  \put(262,49){$-l_c$}
  \put(262,66){$0$}
  \put(262,83){$l_c$}
  \put(262,100){$2l_c$}
  \put(80,114){$\rho_c$}
  \put(72,111){\line(1,0){31}}
\end{picture}
  \vspace{0cm}
  \caption{(a) The smooth cylinder. (b) The fuzzy cylinder. Each segment
  occupies the constant area $2\pi a = 2\pi \rho_c l_c$ 
  and adjacent segments are joined without overlap.}
  \label{fig:cyl}
\end{center}
\end{figure}

\vspace{0.3cm}
First of all let us investigate the fuzzy cylinder as a simple example
of the fuzzy surfaces which are axially symmetric around the $x^3$ 
direction\cite{Hya}.
In order to find an explicit representation of matrices for the fuzzy cylinder,
it is useful to note that the smooth cylinder which is extending to the 
$x^3$ direction is algebraically expressed as
\begin{alignat}{3}
  -\infty < x^3 < \infty , \qquad
  (x^1)^2 + (x^2)^2 = \rho_c^2 . \label{eq:cyls}
\end{alignat}
Here the constant $\rho_c$ is the radius of the smooth cylinder 
(see Fig.\;\ref{fig:cyl}(a)).

Let us consider matrix versions of the above equations.
First we diagonalize the hermitian matrix $X^3$ and interpret each element
as a position in the $x^3$ direction.
Then the first condition in (\ref{eq:cyls}) can be translated into 
the condition that the diagonal elements $X^3_{mm}$
increase from $-\infty$ to $\infty$ as the subscripts $m$ do. 
Here the subscripts $m$ run the whole integer and 
this means that the size of the matrices is infinite.
As mentioned before, this reflects the fact that the cylinder is 
the open surface. Next, by simply replacing variables with matrices 
in the second equation of (\ref{eq:cyls}), we obtain the matrix equation
of the form
\begin{alignat}{3}
(X^1)^2 + (X^2)^2 = \rho_c^2 \mathbf{1}_{\infty}, \label{eq:cylcond}
\end{alignat}
where $\mathbf{1}_{\infty}$ is the $\infty \times \infty$ unit matrix.

Now it is easy to see that the above two conditions for the matrices
$X^1$, $X^2$ and $X^3$ are satisfied if we choose the matrices as
\begin{alignat}{3}
  X^1_{mn} &= \frac{1}{2} \rho_c \delta_{m+1,n} 
  + \frac{1}{2} \rho_c \delta_{m,n+1} ,\notag
  \\
  X^2_{mn} &= \frac{i}{2} \rho_c \delta_{m+1,n} 
  - \frac{i}{2} \rho_c \delta_{m,n+1} , \label{eq:cylmat}
  \\[0.1cm]
  X^3_{mn} &= ml_c \delta_{m,n} , \notag
\end{alignat}
where $m,n \in \mathbb{Z}$. 
We will call the fuzzy element which is located at
$X^3_{mm} = ml_c$ the $m$th segment.
The above matrices are constructed so that
the segments are aligned to the $x^3$ direction
with the constant separation $l_c$. It is interesting to notice that
these matrices satisfy the simple algebra of the form
\begin{alignat}{3}
  [X^1,X^2] = 0, \qquad [X^2,X^3] = il_c X^1, \qquad [X^3,X^1] = il_c X^2. 
  \label{eq:cylalg}
\end{alignat}
This algebra contains only one parameter $l_c$
and the radius $\rho_c$ appears from the Casimir operator\cite{Hya,BL}. 

So far we have obtained the matrices (\ref{eq:cylmat}) by referring to
the smooth cylinder. However the equation (\ref{eq:cylcond}) is not
so strong to determine the representation of matrices uniquely.
For example, it is possible to choose other matrices which 
satisfy (\ref{eq:cylcond}) as
\begin{alignat}{3}
  X^1_{mn} &= \frac{1}{2} \rho_c \delta_{m+k,n} 
  + \frac{1}{2} \rho_c \delta_{m,n+k} ,\notag
  \\
  X^2_{mn} &= \frac{i}{2} \rho_c \delta_{m+k,n} 
  - \frac{i}{2} \rho_c \delta_{m,n+k} , \label{eq:cylmat2}
  \\
  X^3_{mn} &= \frac{m}{k}l_c \delta_{m,n} , \notag
\end{alignat}
where $m,n \in \mathbb{Z}$ and $k$ is an arbitrary positive integer. 
The matrices (\ref{eq:cylmat}) are also included as a case of $k=1$. 
Note that the above matrices satisfy the algebra (\ref{eq:cylalg})
regardless of the value of $k$.

Then in order to fix the matrix representation for the fuzzy cylinder,
we introduce the notion of the area for the fuzzy surface. 
As we will confirm later, 
the area $A$ for the fuzzy surface with axial symmetry is defined as
\begin{alignat}{3}
  A = 2\pi \, \text{Tr} \bigg( \sqrt{ -\frac{1}{2} [X^i,X^j]^2 } \bigg) ,
  \label{eq:area}
\end{alignat}
where $i,j = 1,2,3$ and $i$ and $j$ are contracted respectively. 
It is important to remark that the area for the fuzzy surface becomes
nonzero only when the commutation relations for the matrices $X^i$ 
are nontrivial. In other words, the matrices $X^i$ which are simultaneously
diagonalized represent just a set of points and do not make any surface.
The factor $2\pi$ in the area formula will be explained later.

Now by substituting the matrices (\ref{eq:cylmat2}) for the 
formula (\ref{eq:area}), the area can be estimated as 
$A = \sum_m 2\pi \rho_c l_c$, which means that the each segment 
occupies the constant area $k \times (2\pi \rho_c \tfrac{l_c}{k})$.
From this we see that adjacent segments are joined without any overlap 
only when $k=1$ (see Fig.\;\ref{fig:cyl}(b)).
Therefore we conclude that the matrices (\ref{eq:cylmat}) really 
represent the fuzzy cylinder.

For later use we define $2\pi a = 2\pi \rho_c l_c$ which 
represents the fuzziness of the each matrix diagonal element.

\vspace{0.3cm}
\subsection{The fuzzy surface with axial symmetry}

\begin{figure}[tb]
\begin{center}
  \includegraphics[width=9cm,height=4cm,keepaspectratio]{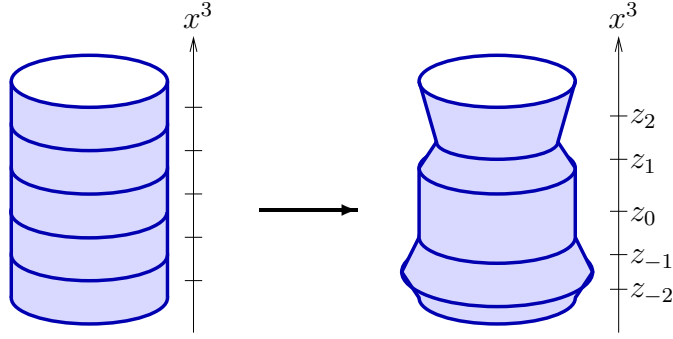}
\begin{picture}(300,0)
  \thicklines
  \put(130,62){\vector(1,0){37}}
  \put(101,131){$x^3$}
  \put(262,131){$x^3$}
  \put(270,29){$z_{-2}$}
  \put(270,42){$z_{-1}$}
  \put(270,58){$z_{0}$}
  \put(270,78){$z_{1}$}
  \put(270,95){$z_{2}$}
\end{picture}
  \vspace{0cm}
  \caption{The general fuzzy surface with axial symmetry 
  is obtained by deforming the fuzzy cylinder.}
  \label{fig:surf}
\end{center}
\end{figure}

\vspace{0.3cm}
In this subsection we generalize the previous arguments to obtain
the matrix representation for a fuzzy surface which has axial symmetry
around the $x^3$ direction.
In order to execute this, let us gaze at the picture of the fuzzy cylinder,
Fig.\;\ref{fig:cyl}(b). There the each value $\rho_c$ 
which appears in the elements $X^{1,2}_{m,m+1}$
or $X^{1,2}_{m+1,m}$ is translated into a position in the radial direction
of the each joint between the $m$th and $(m+1)$th segments,
and the value $X^3_{mm} = ml_c$ is done into a position of the each segment 
in the $x^3$ direction. 
Now it is straightforward to generalize that the fuzzy surface with 
axial symmetry around the $x^3$ direction is represented as
\begin{alignat}{3}
  X^1_{mn} &= \frac{1}{2} \rho_{m+1/2} \delta_{m+1,n} 
  + \frac{1}{2} \rho_{m-1/2} \delta_{m,n+1} ,\notag
  \\
  X^2_{mn} &= \frac{i}{2} \rho_{m+1/2} \delta_{m+1,n} 
  - \frac{i}{2} \rho_{m-1/2} \delta_{m,n+1} , \label{eq:surf}
  \\[0.1cm]
  X^3_{mn} &= z_m \delta_{m,n} , \notag
\end{alignat}
where $m,n$ run an infinite set of integers for the open surface and a finite
set of integers for the closed surface. As is obvious from the explanations
fo far, $z_m$ is interpreted as a position of the $m$th segment
in the $x^3$ direction, and $\rho_{m+1/2}$ is done as a position of 
the joint between the $m$th and $(m+1)$th segments 
in the radial direction (see Fig.\;\ref{fig:surf}). 

It is easy to see that the commutation relations among $X^i$ are evaluated as
\begin{alignat}{3}
  [X^1,X^2]_{mn} &= - \frac{i}{2} \big( \rho_{m+1/2}^2 \!-\! 
  \rho_{m-1/2}^2 \big) \delta_{m,n}, \notag
  \\
  [X^2,X^3]_{mn} &= \frac{i}{2} \rho_{m+1/2} \big( z_{m+1} \!-\! z_m \big) 
  \delta_{m+1,n} 
  + \frac{i}{2} \rho_{m-1/2} \big( z_{m} \!-\! z_{m-1} \big) \delta_{m,n+1} ,
  \label{eq:com}
  \\
  [X^3,X^1]_{mn} &= - \frac{1}{2} \rho_{m+1/2} \big( z_{m+1} \!-\! z_m \big) 
  \delta_{m+1,n} 
  + \frac{1}{2} \rho_{m-1/2} \big( z_{m} \!-\! z_{m-1} \big) \delta_{m,n+1} . 
  \notag
\end{alignat}
Of course, the algebra for the fuzzy cylinder (\ref{eq:cylalg}) 
is reproduced by choosing like
$\rho_{m+1/2} = \rho_c$ and $z_{m+1} \!-\! z_{m} = l_c$ for any $m$.
The commutation relations will become
simple if the differences $z_{m+1} \!-\! z_{m}$ are some constant.

Let ua apply the area formula (\ref{eq:area}) to the fuzzy surface
with axial symmetry.
By substituting the matrices (\ref{eq:surf}) for the formula, 
we obtain the area for the fuzzy surface as
\begin{alignat}{3}
  A &\!=\! 2\pi \!\sum_m\! \sqrt{ \frac{1}{2} \Big\{\! \rho_{m+1/2}^2 
  \big( z_{m+1} \!-\! z_m \big)^2 \!\!+\! \rho_{m-1/2}^2 
  \big( z_{m} \!-\! z_{m-1} \big)^2 \!\Big\}
  \!+\! \frac{1}{4} \big( \rho_{m+1/2}^2 
  \!-\! \rho_{m-1/2}^2 \big)^2 }. \label{eq:area2}
\end{alignat}
Now we take the continuous limit by assuming that the
differences $z_{m+1} \!-\! z_{m}$ are sufficiently small.
In this case the set of $z_m$ becomes a continuous parameter $z$ and
the set of $\rho_{m+1/2}$ is written as a function $z$ as $\rho(z)$. 
Then the first term in the square root is approximated as $\rho^2 dz^2$
and the second term is done as $\frac{1}{4} \{ (\rho^2)'dz \}^2$,
where $'$ is an abbreviation for $\frac{d}{dz}$. 
Furthermore the sum is replaced with the integral on $z$.
After all in the continuous limit, the area formula for the fuzzy surface
(\ref{eq:area2}) reaches to the form
\begin{alignat}{3}
  A &\sim 2\pi \int dz \, \rho \sqrt{ 1 + {\rho'}^2 } . \label{eq:area3}
\end{alignat}
This is precisely coincident with the area for the smooth surface with 
axial symmetry around the $z$ direction. 
Therefore we conclude that the area formula (\ref{eq:area}) 
really gives the area for the fuzzy surface with axial symmetry.
Note that the factor $2\pi$ in the area formula corresponds to 
an integral over the angular direction and this does not arise 
from the trace operation.

Let us investigate the axial symmetry of the fuzzy surface 
in detail. After we rotate the matrices (\ref{eq:surf}) around 
the $x^3$ axis with an angle $\theta$, we obtain 
matrices $\hat{X}^i$ of the forms,
\begin{alignat}{3}
  \hat{X}^1_{mn} &= (\cos\theta X^1 - \sin\theta X^2)_{mn} =
  \frac{1}{2} \rho_{m+1/2} e^{-i\theta} 
  \delta_{m+1,n} + \frac{1}{2} \rho_{m-1/2} e^{i\theta}
  \delta_{m,n+1} ,\notag
  \\
  \hat{X}^2_{mn} &= (\sin\theta X^1 + \cos\theta X^2)_{mn} =
  \frac{i}{2} \rho_{m+1/2} e^{-i\theta} 
  \delta_{m+1,n} - \frac{i}{2} \rho_{m-1/2} e^{i\theta} 
  \delta_{m,n+1} , 
  \\[0.1cm]
  \hat{X}^3_{mn} &= X^3_{mn} = z_m \delta_{m,n} . \notag
\end{alignat}
It is easy to observe that the matrices
$\hat{X}^i$ and $X^i$ hold the three conditions that
$(\hat{X}^1)^2 + (\hat{X}^2)^2 = (X^1)^2 + (X^2)^2$, $\hat{X}^3 = X^3$ 
and $\hat{A} = A$. This means that the matrices $\hat{X}^i$ and $X^i$ represent 
the same fuzzy surface with axial symmetry. The parameter $\theta$ 
corresponds to the `relative' angular direction as we see below.

Furthermore it is important to note that the above global rotation symmetry
around the $x^3$ direction can be elevated to a `local' symmetry as
\begin{alignat}{3}
  \hat{X}^1_{mn} &= \frac{1}{2} \rho_{m+1/2} e^{-i\theta_{m+1/2}} 
  \delta_{m+1,n} + \frac{1}{2} \rho_{m-1/2} e^{i\theta_{m-1/2}} 
  \delta_{m,n+1} ,\notag
  \\
  \hat{X}^2_{mn} &= \frac{i}{2} \rho_{m+1/2} e^{-i\theta_{m+1/2}} 
  \delta_{m+1,n} - \frac{i}{2} \rho_{m-1/2} e^{i\theta_{m-1/2}} 
  \delta_{m,n+1} , \label{eq:surfr}
  \\[0.1cm]
  \hat{X}^3_{mn} &= z_m \delta_{m,n} , \notag
\end{alignat}
where each $\theta_{m+1/2}$ is a real parameter.
It is straightforward to show that these matrices $\hat{X}^i$ 
represent the same fuzzy surface as the matrices $X^i$.
If we explain pictorially,
the matrices $\hat{X}^i$ can be obtained from the matrices $X^i$ by
rotating the $m$th segment with the angle $\phi_m$ where
$\phi_{m+1} \!-\! \phi_{m} = \theta_{m+1/2}$. 

Note that the matrices (\ref{eq:surfr}) 
are also expressed by using a unitary matrix $U$ as
\begin{alignat}{3}
  \hat{X}^i &= U X^i U^{\dagger} ,\qquad
  U_{mn} = e^{i\phi_m} \, \delta_{m,n},
\end{alignat}
where $\phi_m$ are the same as the above.
This implies that the local rotation symmetry around the $x^3$ axis
is identified with $U(1)^{\infty}$ symmetry.

\vspace{0.3cm}
\subsection{The fuzzy sphere}

\vspace{0.3cm}
In this subsection we give self-contained derivation of the matrices 
which represent the fuzzy sphere. Since the fuzzy sphere is axially
symmetric around the $x^3$ direction, we start from the matrices 
(\ref{eq:surf}) and impose the following three conditions.

First, since the smooth sphere with the radius $c$ is algebraically expressed as
\begin{alignat}{3}
  (x^1)^2 + (x^2)^2 + (x^3)^2 = c^2,
\end{alignat}
we impose the condition of the form,
\begin{alignat}{3}
  (X^1)^2 + (X^2)^2 + (X^3)^2 = c^2 \mathbf{1}_{M}, \label{eq:sphcond1}
\end{alignat}
where $\mathbf{1}_M$ is the $M \times M$ unit matrix.
Note that the size of matrices is finite because the sphere 
is the closed surface. In components, the above condition is written as
$\frac{1}{2}(\rho_{m+1/2}^2+\rho_{m-1/2}^2) + z_m^2 = c^2$ for 
$m = 1,2,\cdots,M$ and $\rho_{1/2} = \rho_{M+1/2} = 0$.
Second we require that the each segment of the fuzzy sphere 
occupies the constant area $2\pi a$. 
By referring the area formula (\ref{eq:area2}), 
this condition is expressed as
\begin{alignat}{3}
  \sqrt{ \frac{1}{2} \Big\{\! \rho_{m+1/2}^2 
  \big( z_{m+1} \!-\! z_m \big)^2 \!\!+\! \rho_{m-1/2}^2 
  \big( z_{m} \!-\! z_{m-1} \big)^2 \!\Big\}
  \!+\! \frac{1}{4} \big( \rho_{m+1/2}^2 
  \!-\! \rho_{m-1/2}^2 \big)^2 } = a, \label{eq:sphcond2}
\end{alignat}
where $m=1,2,\cdots,M$. 
Note that in the continuous limit the above equation becomes
\begin{alignat}{3}
  \rho \sqrt{1+{\rho'}^2}dz = c dz = a,
\end{alignat}
for $\rho(z)=\sqrt{c^2 - z^2}$.
This means that the each separation $dz$ between the adjacent segments
is some constant. Then the third condition for the fuzzy sphere is given by
\begin{alignat}{3}
  z_m = \frac{r_s}{M}(2m-M-1). \label{eq:sphcond3}
\end{alignat}
In the continuous limit, the set of $z_m$ becomes
a parameter which range from $-r_s$ to $r_s$, and $r_s$ is thought 
the radius of the sphere.

\begin{figure}[tb]
\begin{center}
  \includegraphics[width=12cm,height=8cm,keepaspectratio]{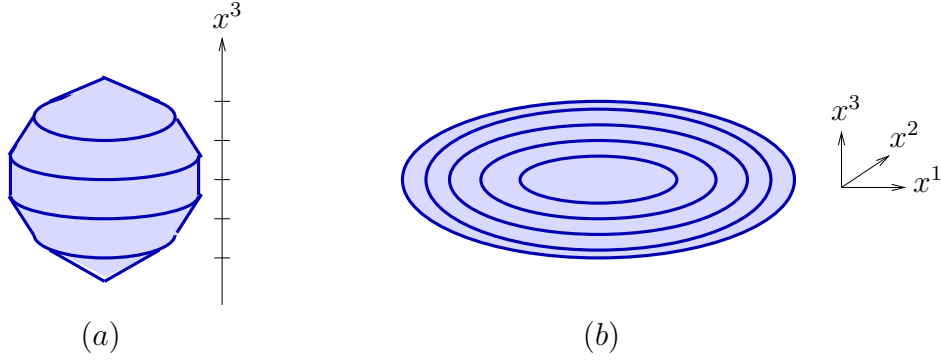}
\begin{picture}(300,0)
  \put(60,120){$x^3$}
  \put(325,58){$x^1$}
  \put(316,75){$x^2$}
  \put(294,85){$x^3$}
  \put(10,0){$(a)$}
  \put(200,0){$(b)$}
\end{picture}
  \vspace{0cm}
  \caption{(a) The fuzzy sphere with $M=5$. (b) The fuzzy plane.}
  \label{fig:sphpla}
\end{center}
\end{figure}

Let us find out the explicit expressions for $\rho_{m+1/2}$ which 
satisfy the conditions (\ref{eq:sphcond1}), (\ref{eq:sphcond2}) 
and (\ref{eq:sphcond3}). From these we obtain the recurrence relations
\begin{alignat}{3}
  \sqrt{4a^2 \!+\! \Big(\frac{2r_s}{M}\Big)^4 \!\!- 4\Big(\frac{2r_s}{M}\Big)^2
  \rho_{m+1/2}^2} &= \sqrt{4a^2 \!+\! \Big(\frac{2r_s}{M}\Big)^4 
  \!\!- 4\Big(\frac{2r_s}{M}\Big)^2 \rho_{m-1/2}^2} 
  \pm 2\Big(\frac{2r_s}{M}\Big)^2.
\end{alignat}
The sign $-$ is for $1 \le m \le [\frac{M}{2}]$ and
the sign $+$ is for $[\frac{M+1}{2}] < m \le M$ since
the condition (\ref{eq:sphcond1}) insists that $\rho_{m+1/2}$ increase
when $z_m < 0$ and decrease when $0 < z_m$.
Then by noting the relations $\rho_{1/2} = \rho_{M+1/2}=0$ and 
$\rho_{5/2}^2 = 2(z_1^2 - z_2^2)$, the above recurrence relations 
can be solved as
\begin{alignat}{3}
  \rho_{m+1/2} = \frac{2r_s}{M} \sqrt{m(M-m)} ,
\end{alignat}
and simultaneously $a$ and $c$ are also determined as
\begin{alignat}{3}
  2\pi a = \frac{4\pi r_s^2}{M} \sqrt{1-\tfrac{1}{M^2}} ,\qquad
  c= r_s \sqrt{1-\tfrac{1}{M^2}}.
\end{alignat}
It is important to note that $2\pi a$ and $c$ become $\tfrac{4\pi r_s^2}{M}$
and $r_s$ respectively in the continuous limit. 
These facts convince us that what we obtained so far precisely describes
the fuzzy sphere (see Fig.\;\ref{fig:sphpla}(a)).

Finally the matrices for the fuzzy sphere are written as
\begin{alignat}{3}
  X^1_{mn} &= \frac{r_s}{M} \sqrt{m(M\!-\!m)} \delta_{m+1,n} 
  + \frac{r_s}{M} \sqrt{(m\!-\!1)(M\!-\!m\!+\!1)} \delta_{m,n+1}
  ,\notag
  \\
  X^2_{mn} &= \frac{ir_s}{M} \sqrt{m(M\!-\!m)} \delta_{m+1,n} 
  - \frac{ir_s}{M} \sqrt{(m\!-\!1)(M\!-\!m\!+\!1)} \delta_{m,n+1} 
  , \label{eq:sphmat}
  \\
  X^3_{mn} &= \frac{r_s}{M} (2m\!-\!M\!-\!1) \delta_{m,n} , \notag
\end{alignat}
where $m,n = 1,2,\cdots,M$.
And the algebra for the fuzzy sphere is given by
\begin{alignat}{3}
  [X^i,X^j] = i\epsilon^{ijk} \frac{2r_s}{M} X^k ,
\end{alignat}
where $i,j,k = 1,2,3$ and $k$ is contracted.

\vspace{0.3cm}
\subsection{The fuzzy plane}

\vspace{0.3cm}
In this subsection we find out the matrices which represent the fuzzy plane.
Since the fuzzy plane has axial symmetry around the $x^3$ direction, 
we start from the matrices (\ref{eq:surf}) and impose the following
two conditions. 

The first one is given by
\begin{alignat}{3}
  X^3_{mn} = z_0 \delta_{m,n}, \label{placond1}
\end{alignat}
where $z_0$ is some constant and $m,n = 1,2,\cdots$. 
The size of matrices is infinite because the fuzzy plane is the open surface.
The above condition means that the fuzzy plane is
located at $z_0$ in the $x^3$ direction.
The second condition is given by
\begin{alignat}{3}
  \frac{1}{2} (\rho_{m+1/2}^2 
  \!-\! \rho_{m-1/2}^2) = a, \label{eq:placond2}
\end{alignat}
with $\rho_{1/2}=0$.
This comes from the requirement that the each segment of the fuzzy
plane occupies the constant area $2\pi a$.
It is easy to see that the above recurrence relations can
be solved as $\rho_{m+1/2} = \sqrt{2am}$ (see Fig.\;\ref{fig:sphpla}(b)).

Then the matrices for the fuzzy plane are represented as
\begin{alignat}{3}
  X^1_{mn} &= \frac{1}{2} \sqrt{2am} \delta_{m+1,n} 
  + \frac{1}{2} \sqrt{2a(m\!-\!1)} \delta_{m,n+1} ,\notag
  \\
  X^2_{mn} &= \frac{i}{2} \sqrt{2am} \delta_{m+1,n} 
  - \frac{i}{2} \sqrt{2a(m\!-\!1)} \delta_{m,n+1} , \label{eq:plamat}
  \\[0.1cm]
  X^3_{mn} &= z_0 \delta_{m,n} , \notag
\end{alignat}
where $m,n = 1,2,\cdots$. And the algebra for the fuzzy plane is given by
\begin{alignat}{3}
  [X^1,X^2] = -ia , \qquad [X^2,X^3] = [X^3,X^1] = 0.
\end{alignat}

Finally it is important to note that the area formula for the fuzzy 
plane becomes
\begin{alignat}{3}
  A = 2\pi \sum_m a = 2\pi \sum_m \rho_m l_m \sim 2\pi \int \rho d\rho,
\end{alignat}
where $\rho_m$ and $l_m$ are defined as 
$\rho_m = \frac{1}{2}(\rho_{m+1/2} + \rho_{m-1/2})$ and 
$l_m = \rho_{m+1/2} - \rho_{m-1/2}$ respectively.
The notation $\sim$ is used to represent the continuous limit.
The above equation precisely represents the area for the smooth plane.
It is also interesting to note that the above equation is also be expressed as
\begin{alignat}{3}
  A = 2\pi i \text{Tr} [X^1,X^2] \sim 2\pi \int \rho d\rho.
\end{alignat}
This relation is often used to show that the D2-brane charge can
be constructed from multiple D0-branes\cite{BFSS,BSS}.

\vspace{1cm}
\section{Effective Actions for the Single D2-brane with Axial Symmetry} 
\label{sec:Effe}

\vspace{0.3cm}
\subsection{Preliminaries}

\vspace{0.3cm}
It is well-known that the bosonic part of the effective action 
for a single D2-brane is described by the abelian Born-Infeld action\cite{Lei}.
In this section 
we explicitly construct the effective actions for the cylindrical, spherical
and planar D2-brane in the background of the flat space-time.

The abelian Born-Infeld action in the background of the flat space-time 
is given by
\begin{alignat}{3}
  S_{\text{D}2} &= - T_2 \int d^3\xi \sqrt{-\det \big(P[\eta]_{\alpha\beta}
  + \lambda F_{\alpha\beta} \big)} , \label{eq:D2}
\end{alignat}
where $T_2 = 1/(2\pi)^2 \ell_s^3 g_s$ is the tension of the D2-brane
and $\ell_s$ and $g_s$ are the string length and the string coupling constant
respectively. And the flat space-time metric is denoted as $\eta_{\mu\nu}$
where $\mu,\nu=0,1,\cdots,9$.
The world-volume coordinates are represented by
$\xi^\alpha (\alpha=0,1,2)$ and $\lambda = 2\pi \ell_s^2$.
Other conventions will be explained later.

The procedures to obtain the effective actions are straightforward, but
we execute without any omission since the results obtained in this section 
will play an important role to justify those obtained in the next section.

\vspace{0.3cm}
\subsection{The effective action for the cylindrical D2-brane}

\vspace{0.3cm}
Let us describe the effective action for the cylindrical D2-brane.
First of all, since the D2-brane is embedded into the target space in the
shape of the cylinder, it is convenient to employ the cylindrical coordinates on
the $(x^1,x^2,x^3)$ part. So the line element of the flat space-time 
is given by
\begin{alignat}{3}
  ds^2 &= -dt^2 + d\rho^2 + \rho^2 d\phi^2 + dz^2 + \sum_{i=4}^9 (dx^i)^2 .
  \label{eq:cylcoo}
\end{alignat}
Then the world-volume coordinates on the cylindrical D2-brane are
chosen as $(t,\phi,z)$
and the D2-brane is embedded into the target space like
$\rho = \rho_c$ and $x^i = 0$. The constant $\rho_c$ is the radius of 
the cylinder. 

Furthermore we add a fluctuation $\hat{\rho}(t,\phi,z)$ 
around $\rho_c$, but neglect other fluctuations around $x^i=0$ for simplicity.
Therefore the scalar fields on the D2-brane are written as 
$\rho = \rho_c + \hat{\rho}$ and $x^i=0$.
Then the pullback metric on the D2-brane is given by
\begin{alignat}{3}
  P[\eta]_{\alpha\beta} = 
  \begin{pmatrix}
    -1+\dot{\hat{\rho}}^2 & \dot{\hat{\rho}}\tilde{\hat{\rho}} 
    & \dot{\hat{\rho}}{\hat{\rho}}' \\
    \dot{\hat{\rho}}\tilde{\hat{\rho}} & \rho^2 + \tilde{\hat{\rho}}^2 
    & \tilde{\hat{\rho}}{\hat{\rho}}' \\
    \dot{\hat{\rho}}{\hat{\rho}}' & \tilde{\hat{\rho}}{\hat{\rho}}' 
    & 1+{\hat{\rho}}'^2
  \end{pmatrix}, \label{eq:pulcyl}
\end{alignat}
where $\alpha,\beta = t,\phi,z$ and we introduced abbreviations
$\dot{\hat{\rho}} = \partial_t \hat{\rho}$, $\tilde{\hat{\rho}} = 
\partial_\phi \hat{\rho}$ and ${\hat{\rho}}' = \partial_z \hat{\rho}$.
The symbol $P[\cdots]$ is used to clarify the pullback operation.
In addition to the scalar fields, 
the abelian gauge field $A_\alpha$ appears from
the massless excitation modes of the open string attached to the D2-brane. 
The field strength is denoted as $F_{\alpha\beta}$.

By substituting the pullback (\ref{eq:pulcyl}) for the action (\ref{eq:D2}),
we obtain the effective action of the form,
\begin{alignat}{3}
  S_{\text{D}2} 
  &= -T_2 \int dt d\phi dz \sqrt{-\det 
  \begin{pmatrix}
    -1+ \dot{\hat{\rho}}^2 & \dot{\hat{\rho}}\tilde{\hat{\rho}} 
    + \lambda F_{t\phi} & \dot{\hat{\rho}}{\hat{\rho}}' + \lambda F_{tz} 
    \\
    \dot{\hat{\rho}}\tilde{\hat{\rho}} - \lambda F_{t\phi} 
    & \rho^2 + \tilde{\hat{\rho}}^2 
    & \tilde{\hat{\rho}}{\hat{\rho}}' + \lambda F_{\phi z} 
    \\
    \dot{\hat{\rho}}{\hat{\rho}}' -\lambda F_{tz} 
    & \tilde{\hat{\rho}}{\hat{\rho}}'
    - \lambda F_{\phi z} & 1 + {\hat{\rho}}'^2 
  \end{pmatrix}} \notag
  \\[0.1cm]
  &= - T_2 \int dt d\phi dz \, \rho
  \sqrt{ 1 + \partial_\alpha \hat{\rho} \partial^\alpha \hat{\rho} 
  + \frac{\lambda^2}{2} F_{\alpha\beta}F^{\alpha\beta} 
  - \frac{\lambda^2}{4} \big(\epsilon^{\alpha\beta\gamma} 
  \partial_\alpha \hat{\rho} \, F_{\beta\gamma} \big)^2 }. 
\end{alignat}
This is the effective action for the cylindrical D2-brane.
The indices $\alpha,\beta$ are raised or lowered by the metric
$ds^2 = -dt^2 + \rho^2 d\phi^2 + dz^2$ and 
$\epsilon^{t\phi z} = 1/\rho$.

In the next section we will construct a single D2-brane from multiple D0-branes.
To be precise, since the D0-brane charge corresponds to
the magnetic flux on the D2-brane, from multiple D0-branes,
we will be able to construct the D2-brane with constant magnetic flux
on the world-volume.
Therefore we must consider the case where the uniform magnetic
flux exists on the D2-brane world-volume.

The uniform magnetic flux on the cylindrical D2-brane
is described by choosing the field strength as 
$F_{\phi z} = \rho_c/b$. And from the quantization condition
of the magnetic flux, 
\begin{alignat}{3}
  \frac{1}{2\pi} \int d\phi dz F_{\phi z} 
  &= \frac{2\pi \rho_c \int dz}{2\pi b} \in \mathbb{Z},
\end{alignat}
the constant $2\pi b$ is interpreted as the area occupied 
per a unit of magnetic flux. Then a fluctuation around this magnetic
flux background is added as
\begin{alignat}{3}
  F_{\phi z} = \frac{\rho_c}{b} + f_{\phi z},
\end{alignat}
where $f_{\phi z}=\partial_\phi a_z \!-\! \partial_z a_\phi$.
We also express the other field strengths by using small letters as 
$F_{t\phi} = f_{t\phi}=\partial_t a_\phi \!-\! \partial_\phi a_t$ 
and $F_{tz} = f_{tz}=\partial_t a_z \!-\! \partial_z a_t$ 
to clarify that these are the fluctuations around trivial backgrounds.

Finally it is natural to consider that 
the fluctuation $\hat{\rho}$ is sufficiently smaller
than the radius $\rho_c$, that is, $\hat{\rho}/\rho_c \ll 1$. 
In this case the effective action for the cylindrical D2-brane 
is estimated as
\begin{alignat}{3}
  S_{\text{D}2} &\!=\! - T_2 \!\int\! dt d\phi dz \, \rho_c
  \sqrt{ 1 + \partial_\alpha \hat{\rho} \partial^\alpha \hat{\rho} 
  + \frac{\lambda^2}{2} F_{\alpha\beta}F^{\alpha\beta} 
  - \frac{\lambda^2}{4} \big(\epsilon^{\alpha\beta\gamma} 
  \partial_\alpha \hat{\rho} \, F_{\beta\gamma} \big)^2 
  + \mathcal{O} \big(\tfrac{\hat{\rho}}{\rho_c} \big) } , \notag
  \\[0.1cm]
  &\!=\! - \frac{T_0}{\lambda} \!\int\! dt dz \, \rho_c
  \Big[ 1 + \big( \!-\! \dot{\hat{\rho}}^2 \!+\! {\hat{\rho}}'^2 \big)
  + \lambda^2 \Big\{ \!-\! \frac{1}{\rho_c^2}{\dot{a}_\phi}^2 
  \!+\! \frac{1}{\rho_c^2} \Big( \frac{\rho_c}{b} 
  \!-\! a'_\phi \Big)^2 \!-\! f_{tz}^2 \Big\} \label{eq:D2cyl}
  \\
  &\qquad\qquad\qquad\quad 
  - \frac{\lambda^2}{\rho_c^2} \Big\{ \dot{\hat{\rho}} 
  \Big( \frac{\rho_c}{b} \!-\! a'_\phi \Big)
  + \hat{\rho}' \dot{a}_\phi \Big\}^2 
  + \mathcal{O} \big(\tfrac{\hat{\rho}}{\rho_c} \big) \Big]^{1/2}. \notag
\end{alignat}
Here the indices $\alpha,\beta$ are raised or lowered by the metric
$ds^2 = -dt^2 + \rho_c^2 d\phi^2 + dz^2$ and 
$\epsilon^{t\phi z} = 1/\rho_c$.
The second line holds if we assume that the fluctuations do not
depend on the angular direction $\phi$.
The above effective action will be reconstructed from the nonabelian 
Born-Infeld action for D0-branes in the next section.

\vspace{0.3cm}
\subsection{The effective action for the spherical D2-brane}

\vspace{0.3cm}
In this subsection we construct the effective action 
for the spherical D2-brane.
Since the D2-brane is embedded into the target space spherically,
we employ the spherical coordinates on
the $(x^1,x^2,x^3)$ part. So the line element of the flat space-time 
is described as
\begin{alignat}{3}
  ds^2 &= -dt^2 + dr^2 + r^2 d\theta^2 + r^2\sin^2\theta 
  d\phi^2 + \sum_{i=4}^9 (dx^i)^2 . \label{eq:metsph}
\end{alignat}
Then the world-volume coordinates on the spherical D2-brane are
chosen as $(t,\theta,\phi)$,
and the D2-brane is embedded into the target space like
$r = r_s$ and $x^i = 0$. The constant $r_s$ is the radius of 
the sphere. 

Furthermore we add a fluctuation $\hat{r}(t,\theta,\phi)$ 
around $r=r_s$ but neglect other fluctuations around $x^i=0$ for simplicity.
Therefore the scalar fields on the D2-brane are written as
$r = r_s + \hat{r}$ and $x^i=0$.
And the pullback metric on the D2-brane is given by
\begin{alignat}{3}
  P[\eta]_{\alpha\beta} = 
  \begin{pmatrix}
    -1+\dot{\hat{r}}^2 & \dot{\hat{r}}{\hat{r}}' & \dot{\hat{r}}\tilde{\hat{r}}
    \\
    \dot{\hat{r}}{\hat{r}}' & r^2 + {\hat{r}}'^2 & {\hat{r}}'\tilde{\hat{r}}
    \\
    \dot{\hat{r}}\tilde{\hat{r}} & {\hat{r}}'\tilde{\hat{r}}
    & r^2\sin^2\theta + \tilde{\hat{r}}^2
  \end{pmatrix}, \label{eq:pulsph}
\end{alignat}
where $\alpha,\beta = t,\theta,\phi$ and we introduced abbreviations
$\dot{\hat{r}} = \partial_t \hat{r}$, ${\hat{r}}' = \partial_\theta \hat{r}$
and $\tilde{\hat{r}} = \partial_\phi \hat{r}$.

By substituting the pullback (\ref{eq:pulsph}) for the action (\ref{eq:D2}),
we obtain the effective action of the form,
\begin{alignat}{3}
  S_{\text{D}2} 
  &= -T_2 \int dt d\theta d\phi \sqrt{-\det 
  \begin{pmatrix}
    -1+\dot{\hat{r}}^2 & \dot{\hat{r}}{\hat{r}}' + \lambda F_{t\theta}
    & \dot{\hat{r}}\tilde{\hat{r}} + \lambda F_{t\phi}
    \\
    \dot{\hat{r}}{\hat{r}}' - \lambda F_{t\theta} 
    & r^2 + {\hat{r}}'^2 & {\hat{r}}'\tilde{\hat{r}} + \lambda F_{\theta\phi}
    \\
    \dot{\hat{r}}\tilde{\hat{r}} - \lambda F_{t\phi} 
    & {\hat{r}}'\tilde{\hat{r}} - \lambda F_{\theta\phi}
    & r^2\sin^2\theta + \tilde{\hat{r}}^2
  \end{pmatrix} } \notag
  \\[0.1cm]
  &= - T_2 \int dt d\theta d\phi \, r^2 \sin\theta
  \sqrt{ 1 + \partial_\alpha \hat{r} \partial^\alpha \hat{r} 
  + \frac{\lambda^2}{2} F_{\alpha\beta}F^{\alpha\beta} 
  - \frac{\lambda^2}{4} \big(\epsilon^{\alpha\beta\gamma} 
  \partial_\alpha \hat{r} \, F_{\beta\gamma} \big)^2 }. 
\end{alignat}
This is the effective action for the spherical D2-brane.
The indices $\alpha,\beta$ are raised or lowered by the metric
$ds^2 = -dt^2 + r^2 d\theta^2 + r^2 \sin^2\theta d\phi^2$ and 
$\epsilon^{t\theta\phi} = 1/r^2\sin\theta$.

The uniform magnetic flux on the spherical D2-brane
is described by choosing the field strength as 
$F_{\theta\phi} = r_s^2\sin\theta/b$. And from the quantization condition
of the magnetic flux, 
\begin{alignat}{3}
  \frac{1}{2\pi} \int d\theta d\phi F_{\theta\phi} 
  &= \frac{4\pi r_s^2}{2\pi b} \in \mathbb{Z},
\end{alignat}
the constant $2\pi b$ is interpreted as the area occupied 
per a unit of magnetic flux. A fluctuation around this magnetic flux background 
is added like
\begin{alignat}{3}
  F_{\theta\phi} = \frac{r_s^2 \sin\theta}{b} + f_{\theta\phi},
\end{alignat}
where $f_{\theta\phi}=\partial_\theta a_\phi \!-\! \partial_\phi a_\theta$.
We also express the other field strengths by using small letters as 
$F_{t\theta} = f_{t\theta}=\partial_t a_\theta \!-\! \partial_\theta a_t$ 
and $F_{t\phi} = f_{t\phi}=\partial_t a_\phi \!-\! \partial_\phi a_t$ 
to clarify that these are the fluctuations around trivial backgrounds.

Finally it is suitable to consider that 
the fluctuation $\hat{r}$ is sufficiently smaller
than the radius $r_s$, that is, $\hat{r}/r_s \ll 1$. 
In this case the effective action for the spherical D2-brane 
is evaluated as
\begin{alignat}{3}
  S_{\text{D}2} &\!=\! - T_2 \!\int\! dt d\theta d\phi \, r_s^2 \sin\theta
  \sqrt{ 1 + \partial_\alpha \hat{r} \partial^\alpha \hat{r} 
  + \frac{\lambda^2}{2} F_{\alpha\beta}F^{\alpha\beta} 
  - \frac{\lambda^2}{4} \big(\epsilon^{\alpha\beta\gamma} 
  \partial_\alpha \hat{r} \, F_{\beta\gamma} \big)^2 
  + \mathcal{O} \big(\tfrac{\hat{r}}{r_s} \big) } , \notag
  \\[0.1cm]
  &\!=\! - \frac{T_0}{\lambda} \!\int\! dt d\theta \, r_s^2 \sin\theta
  \Big[ 1 + \Big\{ \!-\! \dot{\hat{r}}^2 \!+\! \frac{1}{r_s^2} {\hat{r}}'^2 
  \Big\} + \lambda^2 \Big\{ \!-\! \frac{1}{r_s^2}f_{t\theta}^2 
  \!+\! \frac{1}{r_s^4 \sin^2\theta} \Big( \frac{r_s^2\sin\theta}{b} 
  \!+\! a'_\phi \Big)^2 \notag
  \\
  &\qquad\qquad \!-\! \frac{1}{r_s^2 \sin^2\theta} \dot{a}_\phi^2 \Big\} 
  - \frac{\lambda^2}{r_s^4 \sin^2\theta} \Big\{ \dot{\hat{r}} 
  \Big( \frac{r_s^2 \sin\theta}{b} \!+\! a'_\phi \Big)
  - \hat{r}' \dot{a}_\phi \Big\}^2 
  + \mathcal{O} \big(\tfrac{\hat{r}}{r_c} \big) \Big]^{1/2}. \label{eq:D2sph}
\end{alignat}
The indices $\alpha,\beta$ are raised or lowered by the metric
$ds^2 = -dt^2 + r_s^2 d\theta^2 + r_s^2 \sin^2\theta d\phi^2$ and 
$\epsilon^{t\theta\phi} = 1/r_s^2\sin\theta$.
The second equation holds when the fluctuations do not
depend on the angular direction $\phi$.
We will reconstruct the above action from the nonabelian Born-Infeld action
for D0-branes in the next section.

\vspace{0.3cm}
\subsection{The effective action for the planar D2-brane}

\vspace{0.3cm}
Let us describe the effective action for the planar D2-brane.
First of all, since the D2-brane is embedded into the target space in the
shape of the plane, we employ the cylindrical coordinates on
the $(x^1,x^2,x^3)$ part. So the line element of the flat space-time 
is given by (\ref{eq:cylcoo}).
Then the world-volume coordinates on the planar D2-brane are
chosen as $(t,\rho,\phi)$
and the D2-brane is embedded into the target space like
$z = z_0$ and $x^i = 0$. The constant $z_0$ represents the position of 
the plane in the $x^3$ direction.

We also include a fluctuation $\hat{z}(t,\rho,\phi)$ 
around $z=z_0$ but neglect other fluctuations around $x^i=0$ for simplicity.
Then the scalar fields on the D2-brane are written as 
$z = z_0 + \hat{z}$ and $x^i=0$.
The pullback metric on the D2-brane is given by
\begin{alignat}{3}
  P[\eta]_{\alpha\beta} = 
  \begin{pmatrix}
    -1+\dot{\hat{z}}^2 & \dot{\hat{z}}{\hat{z}}'
    & \dot{\hat{z}}\tilde{\hat{z}} 
    \\
    \dot{\hat{z}}{\hat{z}}' & 1 + {\hat{z}}'^2
    & {\hat{z}}'\tilde{\hat{z}} 
    \\
    \dot{\hat{z}}\tilde{\hat{z}} & {\hat{z}}'\tilde{\hat{z}} 
    & \rho^2 + \tilde{\hat{z}}^2 
  \end{pmatrix}, \label{eq:pulpla}
\end{alignat}
where $\alpha,\beta = t,\rho,\phi$ and we introduced abbreviations
$\dot{\hat{z}} = \partial_t \hat{z}$, ${\hat{z}}' = \partial_\rho \hat{z}$ 
and $\tilde{\hat{z}} = \partial_\phi \hat{z}$.

By substituting the pullback (\ref{eq:pulpla}) for the action (\ref{eq:D2}),
we obtain the effective action of the form,
\begin{alignat}{3}
  S_{\text{D}2} 
  &= -T_2 \int dt d\rho d\phi \sqrt{-\det 
  \begin{pmatrix}
    -1+\dot{\hat{z}}^2 & \dot{\hat{z}}{\hat{z}}' + \lambda F_{t\rho}
    & \dot{\hat{z}}\tilde{\hat{z}} + \lambda F_{t\phi} 
    \\
    \dot{\hat{z}}{\hat{z}}' - \lambda F_{t\rho} & 1 + {\hat{z}}'^2
    & {\hat{z}}'\tilde{\hat{z}} + \lambda F_{\rho\phi} 
    \\
    \dot{\hat{z}}\tilde{\hat{z}} - \lambda F_{t\phi} 
    & {\hat{z}}'\tilde{\hat{z}} - \lambda F_{\rho\phi} 
    & \rho^2 + \tilde{\hat{z}}^2 
  \end{pmatrix}} \notag
  \\[0.1cm]
  &= - T_2 \int dt d\rho d\phi \, \rho
  \sqrt{ 1 + \partial_\alpha \hat{z} \partial^\alpha \hat{z} 
  + \frac{\lambda^2}{2} F_{\alpha\beta}F^{\alpha\beta} 
  - \frac{\lambda^2}{4} \big(\epsilon^{\alpha\beta\gamma} 
  \partial_\alpha \hat{z} \, F_{\beta\gamma} \big)^2 }. 
\end{alignat}
This is the effective action for the planar D2-brane.
The indices $\alpha,\beta$ are raised or lowered by the metric
$ds^2 = -dt^2 + d\rho^2 + \rho^2 d\phi^2$ and 
$\epsilon^{t\rho\phi} = 1/\rho$.

The uniform magnetic flux on the planar D2-brane
is described by choosing the field strength like 
$F_{\rho\phi} = \rho/b$. And from the quantization condition 
of the magnetic flux,
\begin{alignat}{3}
  \frac{1}{2\pi} \int d\rho d\phi F_{\rho\phi} 
  &= \frac{2\pi \int d\rho \rho}{2\pi b} \in \mathbb{Z},
\end{alignat}
the constant $2\pi b$ is interpreted as the area occupied 
per a unit of magnetic flux. Then a fluctuation around this magnetic 
flux background is added as
\begin{alignat}{3}
  F_{\rho\phi} = \frac{\rho}{b} + f_{\rho\phi},
\end{alignat}
where $f_{\rho\phi}=\partial_\rho a_\phi \!-\! \partial_\phi a_\rho$.
We also express the other field strengths by using small letters as 
$F_{t\rho} = f_{t\rho}=\partial_t a_\rho \!-\! \partial_\rho a_t$ and
$F_{t\phi} = f_{t\phi}=\partial_t a_\phi \!-\! \partial_\phi a_t$ 
to clarify that these are the fluctuations around trivial backgrounds.

Finally in the case where the fluctuations do not depend on 
the angular direction $\phi$,
the effective action for the planar D2-brane is written as
\begin{alignat}{3}
  S_{\text{D}2} &\!=\! - \frac{T_0}{\lambda} \!\int\! dt d\rho \, \rho
  \Big[ 1 + \big( \!-\! \dot{\hat{z}}^2 \!+\! {\hat{z}}'^2 \big)
  + \lambda^2 \Big\{ \!-\! f_{t\rho}^2 
  \!+\! \frac{1}{\rho^2} \Big( \frac{\rho}{b} 
  \!+\! a'_\phi \Big)^2 \!-\! \frac{1}{\rho^2} \dot{a}_\phi^2 
  \Big\} \label{eq:D2pla}
  \\
  &\qquad\qquad\qquad\quad 
  - \frac{\lambda^2}{\rho^2} \Big\{ \dot{\hat{z}} 
  \Big( \frac{\rho}{b} \!+\! a'_\phi \Big)
  - \hat{z}' \dot{a}_\phi \Big\}^2 \Big]^{1/2}. \notag
\end{alignat}
This action will be reconstructed from the nonabelian Born-Infeld action
for multiple D0-branes in the next section.

\vspace{1cm}
\section{Construction of the D2-brane from Multiple D0-branes} \label{sec:Const}

\vspace{0.3cm}
\subsection{Preliminaries}

\vspace{0.3cm}
In this section we construct the D2-brane from multiple D0-branes.
As a preparation for that, in this subsection we investigate the 
properties of the effective action for multiple D0-branes.

It is well-known that the bosonic part of 
the effective action for $M$ D0-branes is described by 
the nonabelian Born-Infeld action\cite{Mye,Tse,Tse2,TR}. 
This action possesses $U(M)$ gauge symmetry
and consists of $U(M)$ gauge field $A_t$ and nine adjoint 
scalars $X^i\, (i=1,\cdots,9)$ which
appear from the massless excitations of the open strings ending on D0-branes.
In the background of the flat space-time the nonabelian Born-Infeld action
for $M$ D0-branes is expressed as
\begin{alignat}{3}
  S_{\text{D}0} &= -T_0 \int dt \, \text{STr} 
  \Big( \sqrt{-P[E+E']_{tt}\det {Q^i}_j} \Big) , \label{eq:D01}
\end{alignat}
where $T_0 = 1/\ell_s g_s$ is the mass of the D0-brane and 
the notation STr represents the symmetrized trace prescription.
The matrices $E_{\mu\nu}$, $E'_{\mu\nu}$ and ${Q^i}_j$ are defined as
\begin{alignat}{3}
  E_{\mu\nu} = \eta_{\mu\nu} ,\qquad E'_{\mu\nu} = \eta_{\mu i}
  ({Q^{-1 i}}_j - {\delta^i}_j) \eta_{j\nu} ,\qquad
  {Q^i}_j = {\delta^i}_j + \frac{i}{\lambda}[X^i,X^j],
\end{alignat}
where $\mu,\nu = 0,1,\cdots,9$ and $i,j=1,2,\cdots,9$.
And $\eta_{\mu\nu}$ represents the 10-dimensional flat space-time metric.
It is important to note that the pullback operations in the above action 
are modified to respect the $U(M)$
gauge symmetry. Namely the partial derivatives should be replaced 
with the covariant derivatives as
\begin{alignat}{3}
  P[E]_{tt} &= -1 + (D_t X^i)^2 , \notag
  \\
  P[E']_{tt} &= D_t X^i {Q^{-1 i}}_j D_t X^j - (D_t X^i)^2 , \label{eq:pul2}
\end{alignat}
where the covariant derivatives are defined as
$D_t X^i = \partial_t X^i + i[A_t, X^i]$.
Then the action (\ref{eq:D01}) is rewritten as
\begin{alignat}{3}
  S_{\text{D}0} &= -T_0 \int dt \, \text{STr} 
  \Big( \sqrt{\det {Q^i}_j - D_t X^i (\det Q) {Q^{-1 i}}_j D_t X^j} \Big). 
  \label{eq:D02}
\end{alignat}
Note that the matrices such as $D_t X^i$ or $[X^i,X^j]$ 
behave as ordinary numbers through the calculations under the symmetrized 
trace operation. So we do not take care of the order of the matrices
at this stage.

Let us consider the case where multiple D0-branes form 
the fuzzy surface in the $(x^1,x^2,x^3)$ space. In order to do this,
we only keep the matrices $A_t$, $X^1$, $X^2$ and $X^3$ in the action 
and neglect the other adjoint scalars $X^4,\cdots,X^9$ for simplicity.
So it is reasonable to regard $i,j=1,2,3$. Then the $3 \times 3$ matrix
${Q^i}_j$ is written as
\begin{alignat}{3}
  {Q^i}_j &= 
  \begin{pmatrix}
    1 & \;\;\;\frac{i}{\lambda}[X^1,X^2] & -\frac{i}{\lambda}[X^3,X^1] \\
    -\frac{i}{\lambda}[X^1,X^2] & 1 & \;\;\;\frac{i}{\lambda}[X^2,X^3] \\
    \;\;\;\frac{i}{\lambda}[X^3,X^1] & -\frac{i}{\lambda}[X^2,X^3] & 1
  \end{pmatrix}.
\end{alignat}
Note that the each element is also the $M \times M$ hermitian matrix.
Now we evaluate the interior of the square root in the action (\ref{eq:D02}).
First the determinant of the $3 \times 3$ matrix ${Q^i}_j$ is given by
\begin{alignat}{3}
  \det {Q^i}_j \stackrel{\text{STr}}{=} 
  1 - \frac{1}{2\lambda^2}[X^i,X^j]^2. \label{eq:det}
\end{alignat}
The notation $\stackrel{\text{STr}}{=}$ is used to emphasize that
the above equation holds under the symmetrized trace prescription. 
Next the cofactor matrix ${\tilde{Q_j}}^i$ of the $3 \times 3$ 
matrix ${Q^i}_j$ is estimated as
\begin{alignat}{3}
  {\tilde{Q_j}}^i = (\det Q) {Q^{-1 i}}_j \stackrel{\text{STr}}{=}
  \delta_j^i - \frac{i}{\lambda}[X^i,X^j] - \frac{1}{4\lambda^2}
  \epsilon_{ikl}[X^k,X^l]\epsilon_{jmn}[X^m,X^n]. 
\end{alignat}
Then the second term in the square root is calculated as
\begin{alignat}{3}
  D_t X^i {\tilde{Q_j}}^i D_t X^j \stackrel{\text{STr}}{=}
  (D_t X^i)^2 - \frac{1}{4 \lambda^2}
  \big( \epsilon_{ijk} D_t X^i [X^j,X^k] \big)^2, \label{eq:cofa}
\end{alignat}
where $\epsilon_{123} = 1$.

Finally substituting the relations (\ref{eq:det}) and (\ref{eq:cofa}) for the
action (\ref{eq:D02}), we obtain the effective action of the form,
\begin{alignat}{3}
  S_{\text{D}0} &= -T_0 \!\int\! dt \text{Tr} \sqrt{ 1 - (D_t X^i)^2
  - \frac{1}{2\lambda^2}[X^i,X^j]^2 + \frac{1}{4 \lambda^2}
  \big( \epsilon_{ijk} \big\{D_t X^i, [X^j,X^k] \big\} \big)^2 }. \label{eq:D0}
\end{alignat}
Here we replaced STr with the ordinary Tr operation by introducing 
$\{A,B\}=\tfrac{1}{2}(AB+BA)$ in the last term in the square root. 
It should be mentioned that this manipulation is
slightly different from the symmetrized trace operation since we symmetrized
not $( \epsilon_{ijk} \big\{D_t X^i, [X^j,X^k] \big\} )^2$ but
$(\epsilon_{ijk} \big\{D_t X^i, [X^j,X^k] \big\})$.
In the following subsections, we will evaluate the above action in the 
background of the fuzzy cylinder, sphere and plane and show that 
the actions obtained there precisely coincides with 
those obtained in the previous section.
In order to achieve this, we need to modify the 
symmetrized trace operation as stated above.

\vspace{0.3cm}
\subsection{The effective action for the fuzzy cylinder}

\vspace{0.3cm}
\subsubsection{The identification of gauge fluctuations}

\vspace{0.3cm}
As we have discussed in the section \ref{sec:Comm}, 
the fuzzy cylinder is described by the 
matrices (\ref{eq:cylmat}). In this subsection we consider fluctuations
around the fuzzy cylinder and identify the gauge degrees of freedom.
First of all we choose the matrices as
\begin{alignat}{3}
  X^1_{mn} &= \frac{1}{2} \rho_c e^{il_c a_z} \Big|_{m+1/2} \delta_{m+1,n} 
  + \frac{1}{2} \rho_c e^{-il_c a_z} \Big|_{m-1/2} \delta_{m,n+1} , \notag
  \\
  X^2_{mn} &= \frac{i}{2} \rho_c e^{il_c a_z} \Big|_{m+1/2} \delta_{m+1,n} 
  - \frac{i}{2} \rho_c e^{-il_c a_z} \Big|_{m-1/2} \delta_{m,n+1} , 
  \label{eq:matcylg}
  \\
  X^3_{mn} &= \big(ml_c - l_c a_\phi \big) \Big|_m \delta_{m,n} 
  \;,\qquad A_{t\,mn} = a_t \Big|_m \delta_{m,n} . \notag
\end{alignat}
Here we employed the notation $f|_{m+1/2}$ which is equivalent to
$f_{m+1/2}$. For example the notation $\rho_c \exp (il_c a_z)|_{m+1/2}$ 
represents $\rho_c \exp (il_c a_{z\, m+1/2})$. It should be mentioned that
the fluctuations $a_z|_{m+1/2}$, $a_\phi|_m$ and $a_t|_m$
depend on the time $t$.

Now let us calculate the interior of the square root 
in the action (\ref{eq:D0}) step by step. 
First the covariant derivatives are given by
\begin{alignat}{3}
  (D_t X^1)_{mn} &= \frac{ia}{2} f_{tz}
  e^{il_c a_z} \Big|_{m+1/2} \delta_{m+1,n} - \frac{ia}{2} f_{tz}
  e^{-il_c a_z} \Big|_{m-1/2} \delta_{m,n+1} , \notag
  \\
  (D_t X^2)_{mn} &= - \frac{a}{2} f_{tz}
  e^{il_c a_z} \Big|_{m+1/2} \delta_{m+1,n} - \frac{a}{2} f_{tz}
  e^{-il_c a_z} \Big|_{m-1/2} \delta_{m,n+1} ,
  \\
  (D_t X^3)_{mn} &= - l_c \dot{a}_\phi \Big|_m \delta_{m,n} ,\notag
\end{alignat}
where $f_{tz}|_{m+1/2}$ is defined as 
$f_{tz}|_{m+1/2} = \dot{a}_{z}|_{m+1/2} - (a_t|_{m+1} - a_t |_m)/l_c$. 
We also used the relation $2\pi \rho_c l_c = 2\pi a$, 
which represents the fuzziness of the D0-brane. 
From these the second term in the square root is evaluated as
\begin{alignat}{3}
  {(D_t X^i)^2}_{mn} &= a^2 \Big( f_{tz}^2 
  + \frac{1}{\rho_c^2} \dot{a}_\phi^2 \Big) \Big|_m \delta_{m,n} , 
  \label{eq:cylcovg}
\end{alignat}
where $f_{tz}^2|_m$ is defined as $f_{tz}^2|_m = \frac{1}{2}
(f_{tz}^2|_{m+1/2} + f_{tz}^2|_{m-1/2})$.

Second the commutators are given by
\begin{alignat}{3}
  \frac{1}{\lambda}[X^1,X^2]_{mn} &= 0 , \notag
  \\
  \frac{1}{\lambda}[X^2,X^3]_{mn} &= \frac{ia^2}{2\lambda\rho_c} 
  \Big(\frac{\rho_c}{a} - a'_\phi \Big) e^{il_c a_z} 
  \Big|_{m+1/2} \delta_{m+1,n} \notag
  \\ 
  &\quad + \frac{ia^2}{2\lambda\rho_c} 
  \Big(\frac{\rho_c}{a} - a'_\phi \Big) e^{-il_c a_z} 
  \Big|_{m-1/2} \delta_{m,n+1} , 
  \\
  \frac{1}{\lambda}[X^3,X^1]_{mn} &= - \frac{a^2}{2\lambda\rho_c} 
  \Big(\frac{\rho_c}{a} - a'_\phi \Big) e^{il_c a_z} 
  \Big|_{m+1/2} \delta_{m+1,n} \notag
  \\
  &\quad + \frac{a^2}{2\lambda\rho_c} 
  \Big(\frac{\rho_c}{a} - a'_\phi \Big) e^{-il_c a_z} 
  \Big|_{m-1/2} \delta_{m,n+1} , \notag
\end{alignat}
where $a'_\phi|_{m+1/2}$ is defined as 
$a'_\phi|_{m+1/2} = (a_\phi|_{m+1} - a_\phi |_m)/l_c$.
From these the third term in the square root is written as
\begin{alignat}{3}
  \Big( \frac{1}{2\lambda^2}[X^i,X^j]^2 \Big)_{mn} 
  &= - \frac{a^4}{\lambda^2\rho_c^2} \Big(\frac{\rho_c}{a} - a'_\phi \Big)^2 
  \Big|_m \delta_{m,n} , \label{eq:cylcommg}
\end{alignat}
where $(\frac{\rho_c}{a} - a'_\phi)^2|_m$ is defined as
$(\frac{\rho_c}{a} - a'_\phi)^2|_m = \frac{1}{2} \{(\frac{\rho_c}{a} 
- a'_\phi)^2|_{m+1/2} + (\frac{\rho_c}{a} - a'_\phi)^2|_{m-1/2} \}$.

Third the matrix
$\frac{\epsilon_{ijk}}{2\lambda} \{D_t X^i,[X^j,X^k]\}$ is calculated as
\begin{alignat}{3}
  \Big(\frac{\epsilon_{ijk}}{2\lambda} \big\{D_t X^i,[X^j,X^k]\big\} 
  \Big)_{mn} &= 0.
\end{alignat}
From this it is obvious that the fourth term in the square root is zero.

It is important to remark that both (\ref{eq:cylcovg}) and (\ref{eq:cylcommg})
are proportional to the unit matrix. This means that the square root part
in the action (\ref{eq:D0}) is also proportional to the unit matrix
and the trace operation becomes just the sum of the diagonal elements.
Then we obtain the action of the form
\begin{alignat}{3}
  S_{\text{D}0} &= -T_0 \int dt \sum_m \frac{\rho_c l_c}{a}
  \Big[ 1 + a^2 \Big\{\! - \frac{1}{\rho_c^2} \dot{a}_\phi^2 
  + \frac{a^2}{\lambda^2\rho_c^2} \Big(\frac{\rho_c}{a} - a'_\phi \Big)^2 
  - f_{tz}^2  \Big\} \Big]^{1/2} \Big|_m .
\end{alignat}
This is the effective action for the fuzzy cylinder with only gauge fields.

Let us consider the continuous limit of the above action.
When the separation parameter $l_c$ is sufficiently small, 
it is possible to replace $\sum_m l_c$
with $\int dz$. Then the above action reaches to the form
\begin{alignat}{3}
  S_{\text{D}0} &= -\frac{T_0}{a} \int dt dz \, \rho_c
  \Big[ 1 + a^2 \Big\{\! - \frac{1}{\rho_c^2} \dot{a}_\phi^2 
  + \frac{a^2}{\lambda^2\rho_c^2} \Big(\frac{\rho_c}{a} - a'_\phi \Big)^2 
  - f_{tz}^2  \Big\} \Big]^{1/2} .
\end{alignat}
It is easy to see that this action precisely coincides with the 
action (\ref{eq:D2cyl}) with $\hat{\rho} = 0$ in the case of 
$a = b = \lambda$.
Therefore we conclude that the matrices (\ref{eq:matcylg}) correctly describe
the gauge fluctuations on the D2-brane under the condition of 
$a = b = \lambda$.

\vspace{0.3cm}
\subsubsection{The identification of a scalar fluctuation}

\vspace{0.3cm}
In the previous subsection the action (\ref{eq:D2cyl}) with $\hat{\rho}=0$
is realized from the nonabelian Born-Infeld action for D0-branes.
Here we discuss the appearance of a scalar fluctuation from that action.
The starting point is the matrices of the form
\begin{alignat}{3}
  X^1_{mn} &= \frac{1}{2} (\rho_c+\hat{\rho}) \Big|_{m+1/2} \delta_{m+1,n} 
  + \frac{1}{2} (\rho_c+\hat{\rho}) \Big|_{m-1/2} \delta_{m,n+1} , \notag
  \\
  X^2_{mn} &= \frac{i}{2} (\rho_c+\hat{\rho}) \Big|_{m+1/2} \delta_{m+1,n} 
  - \frac{i}{2} (\rho_c+\hat{\rho}) \Big|_{m-1/2} \delta_{m,n+1} ,
  \label{eq:matcyls}
  \\[0.1cm]
  X^3_{mn} &= ml_c \delta_{m,n} ,\qquad A_{t\,mn} = 0 . \notag
\end{alignat}
Note that the fluctuation $\hat{\rho}|_{m+1/2}$ depends on the time $t$.
We neglect the gauge fluctuations discussed in the previous subsection
in order to concentrate on the scalar fluctuation.

Let us evaluate the interior of the square root in the action (\ref{eq:D0})
as before. First the covariant derivatives 
are given by
\begin{alignat}{3}
  (D_t X^1)_{mn} &= \frac{1}{2} \dot{\hat{\rho}} \Big|_{m+1/2} \delta_{m+1,n} 
  + \frac{1}{2} \dot{\hat{\rho}} \Big|_{m-1/2} \delta_{m,n+1} , \notag
  \\
  (D_t X^2)_{mn} &= \frac{i}{2} \dot{\hat{\rho}} \Big|_{m+1/2} \delta_{m+1,n} 
  - \frac{i}{2} \dot{\hat{\rho}} \Big|_{m-1/2} \delta_{m,n+1} , 
  \\[0.1cm]
  (D_t X^3)_{mn} &= 0 . \notag
\end{alignat}
From these the second term in the square root is written as
\begin{alignat}{3}
  {(D_t X^i)^2}_{mn} &= \dot{\hat{\rho}}^2 \Big|_m \delta_{m,n} , 
  \label{eq:cylcovs}
\end{alignat}
where the right hand side is defined as $\dot{\hat{\rho}}^2|_m = \frac{1}{2} 
(\dot{\hat{\rho}}^2|_{m+1/2} + \dot{\hat{\rho}}^2|_{m-1/2})$.

Next the commutators are calculated as
\begin{alignat}{3}
  \frac{1}{\lambda}[X^1,X^2]_{mn} &= \Big(\! - \frac{ia}{\lambda} \hat{\rho}'
  + \mathcal{O} \big( \tfrac{\hat{\rho}}{\rho_c} \big) \Big) 
  \Big|_m \delta_{m,n} , \notag
  \\
  \frac{1}{\lambda}[X^2,X^3]_{mn} &= \Big( \frac{ia}{2\lambda} 
  + \mathcal{O}(\tfrac{\hat{\rho}}{\rho_c}) \Big)
  \Big|_{m+1/2} \delta_{m+1,n} + \Big( \frac{ia}{2\lambda} 
  + \mathcal{O}(\tfrac{\hat{\rho}}{\rho_c}) \Big) 
  \Big|_{m-1/2} \delta_{m,n+1} , 
  \\
  \frac{1}{\lambda}[X^3,X^1]_{mn} &= \Big(\! -\frac{a}{2\lambda} 
  + \mathcal{O}(\tfrac{\hat{\rho}}{\rho_c}) \Big)
  \Big|_{m+1/2} \delta_{m+1,n} + \Big( \frac{a}{2\lambda} 
  + \mathcal{O}(\tfrac{\hat{\rho}}{\rho_c}) \Big)
  \Big|_{m-1/2} \delta_{m,n+1} , \notag
\end{alignat}
where $\hat{\rho}'|_m$ is defined as
$\hat{\rho}'|_m = (\hat{\rho}|_{m+1/2}\!-\!\hat{\rho}|_{m-1/2})/l_c$ 
and we used the relation $2\pi \rho_c l_c = 2\pi a$.
Then the third term in the square root becomes
\begin{alignat}{3}
  \Big(\frac{1}{2\lambda^2}[X^i,X^j]^2 \Big)_{mn} &= \Big(
  - \frac{a^2}{\lambda^2} - \frac{a^2}{\lambda^2} \hat{\rho}'^2 
  + \mathcal{O}(\tfrac{\hat{\rho}}{\rho_c}) \Big) \Big|_m \delta_{m,n} . 
  \label{eq:cylcomms}
\end{alignat}

Last the matrix $\frac{\epsilon_{ijk}}{2\lambda} \{D_t X^i,[X^j,X^k]\}$ is 
calculated as
\begin{alignat}{3}
  \Big(\frac{\epsilon_{ijk}}{2\lambda} \big\{D_t X^i,[X^j,X^k]\big\} \Big)_{mn} 
  &= \Big( \frac{ia}{\lambda} \dot{\hat{\rho}} 
  + \mathcal{O}(\tfrac{\hat{\rho}}{\rho_c}) \Big) \Big|_m \delta_{m,n},
\end{alignat}
where $\dot{\hat{\rho}}|_m$ is defined as
$\dot{\hat{\rho}}|_m = \frac{1}{2} (\dot{\hat{\rho}}|_{m+1/2}
+ \dot{\hat{\rho}}|_{m-1/2} )$.
From this the fourth term in the square root is given by
\begin{alignat}{3}
  \Big(\frac{\epsilon_{ijk}}{2\lambda} \big\{D_t X^i,[X^j,X^k]\big\} 
  \Big)_{mn}^2 &= \Big( -\frac{a^2}{\lambda^2} \dot{\hat{\rho}}^2 
  + \mathcal{O}(\tfrac{\hat{\rho}}{\rho_c})
  \Big) \Big|_m \delta_{m,n}. \label{eq:cylpros}
\end{alignat}
This term does not appear in the previous subsection.

It is important to note that the terms (\ref{eq:cylcovs}), 
(\ref{eq:cylcomms}) and (\ref{eq:cylpros}) are proportional to the unit matrix.
This means that the square root part in the action (\ref{eq:D0}) is also 
proportional to the unit matrix and the trace operation becomes just the 
sum of the diagonal elements. Then we obtain the action of the form
\begin{alignat}{3}
  S_{\text{D}0} &= - T_0 \int dt \sum_m \frac{\rho_c l_c}{a}
  \Big[ 1 + \Big(\! - \dot{\hat{\rho}}^2 + \frac{a^2}{\lambda^2} 
  \hat{\rho}'^2 \Big) + \frac{a^2}{\lambda^2} 
  - \frac{a^2}{\lambda^2} \dot{\hat{\rho}}^2 
  + \mathcal{O}(\tfrac{\hat{\rho}}{\rho_c}) \Big]^{1/2} \Big|_m .
\end{alignat}
This is the effective action for the fuzzy cylinder with only the scalar field.

Let us consider the continuous limit of the above action.
When the separation parameter $l_c$ is sufficiently small, 
it is possible to replace $\sum_m l_c$
with $\int dz$. Then the above action reaches to the form
\begin{alignat}{3}
  S_{\text{D}0} &= - \frac{T_0}{a} \int dt dz \, \rho_c
  \Big[ 1 + \Big(\! - \dot{\hat{\rho}}^2 + \frac{a^2}{\lambda^2} 
  \hat{\rho}'^2 \Big) + \frac{a^2}{\lambda^2} 
  - \frac{a^2}{\lambda^2} \dot{\hat{\rho}}^2 
  + \mathcal{O}(\tfrac{\hat{\rho}}{\rho_c}) \Big]^{1/2} .
\end{alignat}
It is easy to see that this action precisely coincides with the 
action (\ref{eq:D2cyl}) with only the fluctuation $\hat{\rho}$, in the case of 
$a = b = \lambda$.
In other words, the matrices (\ref{eq:matcyls}) correctly describe
the scalar fluctuation on the D2-brane under the condition of 
$a = b = \lambda$.

\vspace{0.3cm}
\subsubsection{The effective action for the fuzzy cylinder}

\vspace{0.3cm}
In this subsection we construct the effective action for the fuzzy cylinder by
combining the results obtained so far.
First of all we choose the matrices for the fuzzy cylinder 
with the gauge and scalar fluctuations as
\begin{alignat}{3}
  X^1_{mn} &= \frac{1}{2} (\rho_c+\hat{\rho}) 
  e^{il_c a_z} \Big|_{m+1/2} \delta_{m+1,n} 
  + \frac{1}{2} (\rho_c+\hat{\rho}) 
  e^{-il_c a_z} \Big|_{m-1/2} \delta_{m,n+1} , \notag
  \\
  X^2_{mn} &= \frac{i}{2} (\rho_c+\hat{\rho}) 
  e^{il_c a_z} \Big|_{m+1/2} \delta_{m+1,n} 
  - \frac{i}{2} (\rho_c+\hat{\rho}) 
  e^{-il_c a_z} \Big|_{m-1/2} \delta_{m,n+1} , \label{eq:matcyl}
  \\
  X^3_{mn} &= \big(ml_c - l_c a_\phi \big) \Big|_m \delta_{m,n} 
  ,\qquad A_{t\,mn} = a_t \Big|_m \delta_{m,n} . \notag
\end{alignat}
Here the gauge fluctuations $a_t|_m$, $a_\phi|_m$, $a_z|_{m+1/2}$ and
the scalar fluctuation $\hat{\rho}|_{m+1/2}$ depend on the time $t$.

Now let us calculate the interior of the square root 
in the action (\ref{eq:D0}) step by step. 
First the covariant derivatives are given by
\begin{alignat}{3}
  (D_t X^1)_{mn} &= \frac{1}{2} \Big( \dot{\hat{\rho}} + ia f_{tz} 
  + \mathcal{O}(\tfrac{\hat{\rho}}{\rho_c}) \Big)
  e^{il_c a_z} \Big|_{m+1/2} \delta_{m+1,n} \notag
  \\
  &\quad + \frac{1}{2} \Big( \dot{\hat{\rho}} - ia f_{tz}
  + \mathcal{O}(\tfrac{\hat{\rho}}{\rho_c}) \Big)
  e^{-il_c a_z} \Big|_{m-1/2} \delta_{m,n+1} , \notag
  \\
  (D_t X^2)_{mn} &= \frac{i}{2} \Big( \dot{\hat{\rho}} + ia f_{tz} 
  + \mathcal{O}(\tfrac{\hat{\rho}}{\rho_c}) \Big)
  e^{il_c a_z} \Big|_{m+1/2} \delta_{m+1,n} 
  \\
  &\quad - \frac{i}{2} \Big( \dot{\hat{\rho}} - ia f_{tz}
  + \mathcal{O}(\tfrac{\hat{\rho}}{\rho_c}) \Big)
  e^{-il_c a_z} \Big|_{m-1/2} \delta_{m,n+1} , \notag
  \\
  (D_t X^3)_{mn} &= - l_c \dot{a}_\phi \Big|_m \delta_{m,n} , \notag
\end{alignat}
where $f_{tz}|_{m+1/2}$ is defined as 
$f_{tz}|_{m+1/2} = \dot{a}_{z}|_{m+1/2} - (a_t|_{m+1} - a_t |_m)/l_c$. 
We also used the relation $2\pi \rho_c l_c = 2\pi a$, 
which means that the area occupied per the D0-brane is $2\pi a$. 
From these the second term in the square root is evaluated as
\begin{alignat}{3}
  {(D_t X^i)^2}_{mn} &= \Big\{ \dot{\hat{\rho}}^2 + a^2 \Big( f_{tz}^2 
  + \frac{1}{\rho_c^2} \dot{a}_\phi^2 \Big) 
  + \mathcal{O}(\tfrac{\hat{\rho}}{\rho_c}) \Big\} \Big|_m \delta_{m,n} ,
  \label{eq:cylcov}
\end{alignat}
where $f_{tz}^2|_m$ and $\dot{\hat{\rho}}^2|_m$ are defined as 
$f_{tz}^2|_m = \frac{1}{2} (f_{tz}^2|_{m+1/2} + f_{tz}^2|_{m-1/2})$ and 
$\dot{\hat{\rho}}^2|_m = \frac{1}{2} (\dot{\hat{\rho}}^2|_{m+1/2} 
+ \dot{\hat{\rho}}^2|_{m-1/2})$ respectively.

Second the commutators are given by
\begin{alignat}{3}
  \frac{1}{\lambda}[X^1,X^2]_{mn} &= \Big(\! -\frac{ia}{\lambda} 
  \hat{\rho}' + \mathcal{O}(\tfrac{\hat{\rho}}{\rho_c}) \Big) 
  \Big|_m \delta_{m,n} , \notag
  \\
  \frac{1}{\lambda}[X^2,X^3]_{mn} &= \frac{i}{2}
  \Big\{ \frac{a^2}{\lambda\rho_c} \Big(\frac{\rho_c}{a} - a'_\phi \Big) 
  + \mathcal{O}(\tfrac{\hat{\rho}}{\rho_c}) \Big\} e^{il_c a_z} 
  \Big|_{m+1/2} \delta_{m+1,n} \notag
  \\ 
  &\quad + \frac{i}{2} \Big\{ \frac{a^2}{\lambda\rho_c} 
  \Big(\frac{\rho_c}{a} - a'_\phi \Big) 
  + \mathcal{O}(\tfrac{\hat{\rho}}{\rho_c}) \Big\} e^{-il_c a_z} 
  \Big|_{m-1/2} \delta_{m,n+1} , 
  \\
  \frac{1}{\lambda}[X^3,X^1]_{mn} &= - \frac{1}{2}
  \Big\{ \frac{a^2}{\lambda\rho_c} \Big(\frac{\rho_c}{a} - a'_\phi \Big) 
  + \mathcal{O}(\tfrac{\hat{\rho}}{\rho_c}) \Big\} e^{il_c a_z} 
  \Big|_{m+1/2} \delta_{m+1,n} \notag
  \\ 
  &\quad + \frac{1}{2} \Big\{ \frac{a^2}{\lambda\rho_c} 
  \Big(\frac{\rho_c}{a} - a'_\phi \Big) 
  + \mathcal{O}(\tfrac{\hat{\rho}}{\rho_c}) \Big\} e^{-il_c a_z} 
  \Big|_{m-1/2} \delta_{m,n+1} , \notag
\end{alignat}
where $a'_\phi|_{m+1/2}$ and $\hat{\rho}'|_m$ are defined as 
$a'_\phi|_{m+1/2} = (a_\phi|_{m+1} - a_\phi |_m)/l_c$ and
$\hat{\rho}'|_m = (\hat{\rho}|_{m+1/2}\!-\!\hat{\rho}|_{m-1/2})/l_c$ 
respectively. From these the third term in the square root is written as
\begin{alignat}{3}
  \Big(\frac{1}{2\lambda^2}[X^i,X^j]^2 \Big)_{mn} &= \Big\{
  - \frac{a^2}{\lambda^2}\hat{\rho}'^2 - \frac{a^4}{\lambda^2\rho_c^2} 
  \Big(\frac{\rho_c}{a} - a'_\phi \Big)^2 
  + \mathcal{O}(\tfrac{\hat{\rho}}{\rho_c}) \Big\} \Big|_m \delta_{m,n} ,
  \label{eq:cylcomm}
\end{alignat}
where $(\frac{\rho_c}{a} - a'_\phi)^2|_m$ is defined as
$(\frac{\rho_c}{a} - a'_\phi)^2|_m = \frac{1}{2} \{(\frac{\rho_c}{a} 
- a'_\phi)^2|_{m+1/2} + (\frac{\rho_c}{a} - a'_\phi)^2|_{m-1/2} \}$.

Third the matrix 
$\frac{\epsilon_{ijk}}{2\lambda} \big\{D_t X^i,[X^j,X^k]\big\}$ is 
estimated as
\begin{alignat}{3}
  \Big(\frac{\epsilon_{ijk}}{2\lambda} \big\{D_t X^i,[X^j,X^k]\big\} \Big)_{mn}
  &= \Big[ \frac{ia^2}{\lambda\rho_c} \Big\{ \dot{\hat{\rho}} 
  \Big(\frac{\rho_c}{a} - a'_\phi \Big)
  + \hat{\rho}' \dot{a}_\phi \Big\} 
  + \mathcal{O}(\tfrac{\hat{\rho}}{\rho_c}) \Big] \Big|_m \delta_{m,n},
\end{alignat}
where $\dot{\hat{\rho}}(\frac{\rho_c}{a} - a'_\phi)|_m$ is defined as
$\dot{\hat{\rho}}(\frac{\rho_c}{a} - a'_\phi)|_m = \frac{1}{2} 
\{ \dot{\hat{\rho}}(\frac{\rho_c}{a} - a'_\phi)|_{m+1/2}
+ \dot{\hat{\rho}}(\frac{\rho_c}{a} - a'_\phi)|_{m-1/2} \}$.
Then the fourth term in the square root is given by
\begin{alignat}{3}
  \Big(\frac{\epsilon_{ijk}}{2\lambda} 
  \big\{D_t X^i,[X^j,X^k]\big\} \Big)_{mn}^2 
  &= \Big[\! -\frac{a^4}{\lambda^2\rho_c^2} \Big\{ \dot{\hat{\rho}} 
  \Big(\frac{\rho_c}{a} - a'_\phi \Big)
  + \hat{\rho}' \dot{a}_\phi \Big\}^2 
  + \mathcal{O}(\tfrac{\hat{\rho}}{\rho_c}) \Big] \Big|_m \delta_{m,n}.
  \label{eq:cylpro}
\end{alignat}

Note that the terms (\ref{eq:cylcov}), 
(\ref{eq:cylcomm}) and (\ref{eq:cylpro}) are proportional to the unit matrix.
This means that the square root part in the action (\ref{eq:D0}) is also 
proportional to the unit matrix and the trace operation becomes just the 
sum of the diagonal elements. Then we obtain the action of the form
\begin{alignat}{3}
  S_{\text{D}0} &= -T_0 \int dt \sum_m \frac{\rho_c l_c}{a}
  \Big[ 1 + \Big(\! - \dot{\hat{\rho}}^2 
  + \frac{a^2}{\lambda^2} \hat{\rho}'^2 \Big) 
  + a^2 \Big\{ - \frac{1}{\rho_c^2} \dot{a}_\phi^2 
  + \frac{a^2}{\lambda^2\rho_c^2} 
  \Big(\frac{\rho_c}{a} - a'_\phi \Big)^2 - f_{tz}^2 \Big\} \notag
  \\
  &\qquad\qquad\qquad\qquad -\frac{a^4}{\lambda^2\rho_c^2} 
  \Big\{ \dot{\hat{\rho}} \Big(\frac{\rho_c}{a} - a'_\phi \Big)
  + \hat{\rho}' \dot{a}_\phi \Big\}^2 
  + \mathcal{O}(\tfrac{\hat{\rho}}{\rho_c}) 
  \Big]^{1/2} \Big|_m . \label{eq:D0cyl}
\end{alignat}
This is the effective action for the fuzzy cylinder.

Let us consider the continuous limit of the above action.
When the separation parameter $l_c$ is sufficiently small, 
it is possible to replace $\sum_m l_c$
with $\int dz$. Then the above action reaches to the form
\begin{alignat}{3}
  S_{\text{D}0} &= - \frac{T_0}{a} \int dt dz \, \rho_c
  \Big[ 1 + \Big(\! - \dot{\hat{\rho}}^2 
  + \frac{a^2}{\lambda^2} \hat{\rho}'^2 \Big) 
  + a^2 \Big\{ - \frac{1}{\rho_c^2} \dot{a}_\phi^2 
  + \frac{a^2}{\lambda^2\rho_c^2} 
  \Big(\frac{\rho_c}{a} - a'_\phi \Big)^2 - f_{tz}^2 \Big\} \notag
  \\
  &\qquad\qquad\qquad\qquad -\frac{a^4}{\lambda^2\rho_c^2} 
  \Big\{ \dot{\hat{\rho}} \Big(\frac{\rho_c}{a} - a'_\phi \Big)
  + \hat{\rho}' \dot{a}_\phi \Big\}^2 
  + \mathcal{O}(\tfrac{\hat{\rho}}{\rho_c})
  \Big]^{1/2} . \label{eq:D0cylc}
\end{alignat}
This action precisely coincides with the 
action (\ref{eq:D2cyl}) in the case of $a = b = \lambda$.
In other words, the matrices (\ref{eq:matcyl}) correctly describe
the gauge and scalar fluctuations on the D2-brane under the 
condition of $a = b = \lambda$.

From the results obtained so far, we conclude that the gauge fluctuations
on the D2-brane world-volume appear from the scalar fluctuations of
D0-branes which are parallel to the fuzzy cylinder and the scalar fluctuation
on the D2-brane does from the scalar fluctuations of
D0-branes which are perpendicular to the fuzzy cylinder.

\vspace{0.3cm}
\subsection{The effective action for the fuzzy sphere}

\vspace{0.3cm}
As we have discussed in the section \ref{sec:Comm}, 
the fuzzy sphere is described by the 
matrices (\ref{eq:sphmat}). In this subsection we consider the fluctuations
around the fuzzy sphere and construct the effective action for it.
First of all we choose the $M \times M$ matrices as
\begin{alignat}{3}
  X^1_{mn} &= \frac{1}{2} (\rho+\hat{\rho}) e^{i(l/r_s)a_\theta}
  \Big|_{m+1/2} \delta_{m+1,n} 
  + \frac{1}{2} (\rho+\hat{\rho}) e^{-i(l/r_s)a_\theta} 
  \Big|_{m-1/2} \delta_{m,n+1} , \notag
  \\
  X^2_{mn} &= \frac{i}{2} (\rho+\hat{\rho}) e^{i(l/r_s)a_\theta} 
  \Big|_{m+1/2} \delta_{m+1,n} 
  - \frac{i}{2} (\rho+\hat{\rho}) e^{-i(l/r_s)a_\theta} 
  \Big|_{m-1/2} \delta_{m,n+1} , \label{eq:matsph}
  \\
  X^3_{mn} &= (z+\hat{z}) \Big|_m \delta_{m,n} ,\qquad 
  A_{t\,mn} = a_t \Big|_m \delta_{m,n} , \notag
\end{alignat}
where $m,n =1,\cdots,M$.
As we observed in the section \ref{sec:Comm}, 
$\rho_{m+1/2}$ and $z_m$ are the background elements 
which form the fuzzy sphere and explicitly written as 
$\rho_{m+1/2} = \tfrac{2r_s}{M}\sqrt{m(M\!-\!m)}$ and 
$z_m = \tfrac{r_s}{M}(2m\!-\!M\!-\!1)$ respectively.
The constant $r_s$ represents the radius of the sphere.
$\hat{\rho}|_{m+1/2}$, $\hat{z}|_m$, $a_\theta|_{m+1/2}$ and $a_t|_m$ are 
the fluctuations around the fuzzy sphere and depend on the time $t$.
Each length $l_{m+1/2}$ is defined so as to satisfy 
$2\pi \rho_{m+1/2}l_{m+1/2} = 2\pi a$ and interpreted 
as an `arc' between the $m$th and
the $(m\!+\!1)$th D0-branes. In particular the relation 
$l_{m+1/2}/r_s \sim d\theta$ holds in the continuous limit.

Now let us evaluate the interior of the square root 
in the action (\ref{eq:D0}) step by step. 
First the covariant derivatives are given by
\begin{alignat}{3}
  (D_t X^1)_{mn} &= \frac{1}{2} \Big( \dot{\hat{\rho}} 
  + \frac{ia}{r_s} f_{t\theta} + \mathcal{O}(\tfrac{\hat{\rho}}{\rho}) \Big) 
  e^{i(l/r_s) a_\theta} \Big|_{m+1/2} \delta_{m+1,n} \notag
  \\
  &\quad + \frac{1}{2} \Big( \dot{\hat{\rho}} - \frac{ia}{r_s} f_{t\theta}
  + \mathcal{O}(\tfrac{\hat{\rho}}{\rho}) \Big) e^{-i(l/r_s) a_\theta} 
  \Big|_{m-1/2} \delta_{m,n+1} , \notag
  \\
  (D_t X^2)_{mn} &= \frac{i}{2} \Big( \dot{\hat{\rho}} 
  + \frac{ia}{r_s} f_{t\theta} + \mathcal{O}(\tfrac{\hat{\rho}}{\rho}) \Big) 
  e^{i(l/r_s) a_\theta} \Big|_{m+1/2} \delta_{m+1,n} 
  \\
  &\quad - \frac{i}{2} \Big( \dot{\hat{\rho}} - \frac{ia}{r_s} f_{t\theta}
  + \mathcal{O}(\tfrac{\hat{\rho}}{\rho}) \Big) e^{-i(l/r_s) a_\theta} 
  \Big|_{m-1/2} \delta_{m,n+1} , \notag
  \\
  (D_t X^3)_{mn} &= \dot{\hat{z}} \Big|_m \delta_{m,n} , \notag
\end{alignat}
where $f_{t\theta}|_{m+1/2}$ is defined as 
$f_{t\theta}|_{m+1/2} = \dot{a}_\theta|_{m+1/2} 
- (a_t|_{m+1} - a_t|_m)/(l_{m+1/2}/r_s)$.
From these the second term in the square root is calculated as
\begin{alignat}{3}
  {(D_t X^i)^2}_{mn} &= \Big( \dot{\hat{\rho}}^2 + \dot{\hat{z}}^2 
  + \frac{a^2}{r_s^2} f_{t\theta}^2 + \mathcal{O}(\tfrac{\hat{\rho}}{\rho}) 
  \Big) \Big|_m \delta_{m,n} ,
\end{alignat}
where $\dot{\hat{\rho}}^2|_m$ and $f_{t\theta}^2|_m$ are defined as
$\dot{\hat{\rho}}^2|_m = \tfrac{1}{2}(\dot{\hat{\rho}}^2|_{m+1/2} +
\dot{\hat{\rho}}^2|_{m-1/2})$ and 
$f_{t\theta}^2|_m = \tfrac{1}{2}(f_{t\theta}^2|_{m+1/2} +
f_{t\theta}^2|_{m-1/2})$ respectively.

Next the commutators are written as
\begin{alignat}{3}
  \frac{1}{\lambda}[X^1,X^2]_{mn} &= \Big(\! - \frac{ia}{\lambda r_s} 
  (\rho'\!+\!\hat{\rho}') + \mathcal{O}(\tfrac{\hat{\rho}}{\rho}) \Big) 
  \Big|_m \delta_{m,n}, \notag
  \\
  \frac{1}{\lambda}[X^2,X^3]_{mn} &= \Big( \frac{ia}{2\lambda r_s} 
  (z'\!+\!\hat{z}') + \mathcal{O}(\tfrac{\hat{\rho}}{\rho}) \Big) 
  e^{i(l/r_s) a_\theta} \Big|_{m+1/2} \delta_{m+1,n} \notag
  \\
  &\quad
  + \Big( \frac{ia}{2\lambda r_s} (z'\!+\!\hat{z}') 
  + \mathcal{O}(\tfrac{\hat{\rho}}{\rho}) \Big) 
  e^{-i(l/r_s) a_\theta} \Big|_{m-1/2} \delta_{m,n+1} , 
  \\
  \frac{1}{\lambda}[X^3,X^1]_{mn} &= \Big(\! - \frac{a}{2\lambda r_s} 
  (z'\!+\!\hat{z}') + \mathcal{O}(\tfrac{\hat{\rho}}{\rho}) \Big) 
  e^{i(l/r_s) a_\theta} \Big|_{m+1/2} \delta_{m+1,n} \notag
  \\
  &\quad
  + \Big( \frac{a}{2\lambda r_s} (z'\!+\!\hat{z}') 
  + \mathcal{O}(\tfrac{\hat{\rho}}{\rho}) \Big) 
  e^{-i(l/r_s) a_\theta} \Big|_{m-1/2} \delta_{m,n+1} , \notag
\end{alignat}
where $\rho'|_m$ and $z'|_{m+1/2}$ are defined as 
$\rho'|_m = (\rho|_{m+1/2} - \rho|_{m-1/2})/(l_{m}/r_s)$ and
$z'|_{m+1/2} = (z|_{m+1} - z|_m)/(l_{m+1/2}/r_s)$ and so on.
Each length $l_m$ is also defined so as to satisfy the relation
$2\pi \rho_m l_m = 2\pi a$, where 
$\rho_m = \tfrac{1}{2}(\rho_{m+1/2}+\rho_{m-1/2})$.
As like the case of $l_{m+1/2}$, $l_m$ is interpreted as 
an `arc' around the $m$th D0-brane and regarded as
$r_s d\theta$ in the continuous limit.
From the above equations the third term in the square root is given by
\begin{alignat}{3}
  \Big(\frac{1}{2\lambda^2}[X^i,X^j]^2 \Big)_{mn} &= 
  \Big(\! - \frac{a^2}{\lambda^2 r_s^2} (\rho'\!+\!\hat{\rho}')^2
  - \frac{a^2}{\lambda^2 r_s^2} (z'\!+\!\hat{z}')^2 
  + \mathcal{O}(\tfrac{\hat{\rho}}{\rho}) \Big) \Big|_m \delta_{m,n} ,
\end{alignat}
where $(z'\!+\!\hat{z}')^2|_m$ is defined as
$(z'\!+\!\hat{z}')^2|_m = \tfrac{1}{2} 
\big\{ (z'\!+\!\hat{z}')^2|_{m+1/2}+(z'\!+\!\hat{z}')^2|_{m-1/2} \big\}$.

Last the matrix
$\frac{\epsilon_{ijk}}{2\lambda} \big\{D_t X^i,[X^j,X^k]\big\}$ becomes
\begin{alignat}{3}
  \Big( \frac{\epsilon_{ijk}}{2\lambda} 
  \big\{D_t X^i,[X^j,X^k]\big\} \Big)_{mn}
  &= \Big(\! - \frac{ia}{\lambda r_s} \big(\dot{\hat{z}} 
  (\rho'\!+\!\hat{\rho}') - \dot{\hat{\rho}} (z'\!+\!\hat{z}') \big) 
  + \mathcal{O}(\tfrac{\hat{\rho}}{\rho}) \Big) \Big|_m \delta_{m,n} ,
\end{alignat}
where $\dot{\hat{\rho}} (z'\!+\!\hat{z}')|_m$ is defined as
$\dot{\hat{\rho}} (z'\!+\!\hat{z}')|_m = \tfrac{1}{2}
\{\dot{\hat{\rho}} (z'\!+\!\hat{z}')|_{m+1/2} + \dot{\hat{\rho}} 
(z'\!+\!\hat{z}')|_{m-1/2}\}$.
From this the fourth term in the square root is written as
\begin{alignat}{3}
  \Big(\frac{\epsilon_{ijk}}{2\lambda} \big\{D_t X^i,[X^j,X^k]\big\} 
  \Big)_{mn}^2 &\!=\! \Big(\! - \frac{a^2}{\lambda^2 r_s^2} 
  \big(\dot{\hat{z}}(\rho'\!+\!\hat{\rho}') 
  - \dot{\hat{\rho}}(z'\!+\!\hat{z}') \big)^2 
  + \mathcal{O}(\tfrac{\hat{\rho}}{\rho}) \Big) \Big|_m \delta_{m,n} .
\end{alignat}

Now we are ready to evaluate the action (\ref{eq:D0}). 
From the calculations
so far, it is easy to see that the trace operation becomes just the 
summation since the the square root part is proportional to 
the unit matrix. Then we obtain the action of the form,
\begin{alignat}{3}
  S_{\text{D}0} &= -T_0 \int dt \sum_m \frac{\rho l}{a} 
  \Big[1 - \Big( \dot{\hat{\rho}}^2 + \dot{\hat{z}}^2 
  + \frac{a^2}{r_s^2} f_{t\theta}^2 \Big) + \Big( \frac{a^2}{\lambda^2 r_s^2}
  (\rho'\!+\!\hat{\rho}')^2 + \frac{a^2}{\lambda^2 \rho_s^2} 
  (z'\!+\!\hat{z}')^2 \Big) \notag
  \\
  &\qquad\qquad\qquad\qquad\quad
  - \frac{a^2}{\lambda^2 r_s^2} \big(\dot{\hat{z}}(\rho'\!+\!\hat{\rho}') 
  - \dot{\hat{\rho}}(z'\!+\!\hat{z}') \big)^2 
  + \mathcal{O}(\tfrac{\hat{\rho}}{\rho}) \Big]^{1/2} \Big|_m . 
  \label{eq:D0sph}
\end{alignat}
This is the effective action for the fuzzy sphere.
At this stage, the correspondence between this action and 
the effective action for the spherical D2-brane (\ref{eq:D2sph})
is not clear.

In order to observe the correspondence, we need to decompose the fluctuations
$\dot{\hat{\rho}}|_m$, $\dot{\hat{z}}|_m$, $\hat{\rho}'|_m$ and 
$\hat{z}'|_m$ as
\begin{alignat}{3}
  \dot{\hat{\rho}} \Big|_m &= \frac{\rho}{r_s} \dot{\hat{r}} \Big|_m 
  + \frac{z}{r_s} \Big( \frac{a}{\rho} \dot{a}_\phi \Big) \Big|_m ,
  \qquad &
  \dot{\hat{z}} \Big|_m &= \frac{z}{r_s}\dot{\hat{r}} \Big|_m 
  - \frac{\rho}{r_s} \Big( \frac{a}{\rho} \dot{a}_\phi \Big) \Big|_m , \notag
  \\[0.1cm]
  \hat{\rho}' \Big|_m &= \frac{\rho}{r_s} \hat{r}' \Big|_m 
  + \frac{z}{r_s} \Big( \frac{a}{\rho} a'_\phi \Big) \Big|_m ,
  \qquad &
  \hat{z}' \Big|_m &= \frac{z}{r_s}\hat{r}' \Big|_m 
  - \frac{\rho}{r_s} \Big( \frac{a}{\rho} a'_\phi \Big) \Big|_m ,
\end{alignat}
where $\dot{\hat{r}}$ or $\hat{r}'$ are the bases perpendicular 
to the fuzzy sphere and $\frac{a}{\rho} \dot{a}_\phi$ 
or $\tfrac{a}{\rho} a'_\phi$ are those along the fuzzy sphere.

Let us consider the continuous limit by taking $M$ sufficiently large.
Then $\rho|_m$ and $l|_m$ are replaced with
$r_s \sin\theta$ and $r_s d\theta$ respectively and $\rho^2|_m + z^2|_m$
becomes $r_s^2$. It is also useful to notice that
\begin{alignat}{3}
  (\rho'+\hat{\rho}')^2+(z'+\hat{z}')^2 &= \hat{r}'^2 
  + \Big( r_s + \frac{a}{r_s \sin\theta} a_\phi' \Big)^2, \notag
  \\
  \dot{\hat{z}}(\rho'\!+\!\hat{\rho}') - \dot{\hat{\rho}} (z'\!+\!\hat{z}') 
  &= \dot{\hat{r}} \Big( r_s + \frac{a}{r_s \sin\theta} a_\phi' \Big)
  - \hat{r}' \Big( \frac{a}{r_s \sin\theta} \dot{a}_\phi \Big) .
\end{alignat}
By applying these relations to the action, it reaches to the form
\begin{alignat}{3}
  S_{\text{D}0} &\!=\! -\frac{T_0}{a} \!\!\int\!\! dt d\theta \, 
  r_s^2 \sin\theta \Big[1 \!+\! \Big(\!\! - \dot{\hat{r}}^2 \!+\! 
  \frac{a^2}{\lambda^2 r_s^2} \hat{r}'^2 \Big) \!+\! a^2 
  \Big\{\!\! - \frac{1}{r_s^2} f_{t\theta}^2 \!+\!
  \frac{a^2}{\lambda^2 r_s^4 \sin^2\theta} \Big( \frac{r_s^2}{a} 
  \sin\theta \!+\! a'_\phi \Big)^2 \notag
  \\
  &\qquad\qquad \!-\! \frac{1}{r_s^2 \sin^2\theta} \dot{a}_\phi^2 \Big\} 
  \!-\! \frac{a^4}{\lambda^2 r_s^4 \sin^2\theta} \Big\{ \dot{\hat{r}} 
  \Big( \frac{r_s^2}{a}\sin\theta \!+\! a_\phi' \Big) \!-\! \hat{r}' 
  \dot{a}_\phi \Big\}^2 \!+ \mathcal{O}(\tfrac{\hat{r}}{r_s}) 
  \Big]^{1/2} . \label{eq:D0sphc}
\end{alignat}
This action precisely coincides with the 
action (\ref{eq:D2sph}) in the case of $a = b = \lambda$.
In other words, the matrices (\ref{eq:matsph}) correctly reproduce
the gauge and scalar fluctuations on the D2-brane under the 
condition of $a = b = \lambda$.

From the results obtained in this subsection, 
we conclude that the gauge fluctuations
on the D2-brane world-volume arise from the scalar fluctuations of
D0-branes which are parallel to the fuzzy sphere and the scalar fluctuation
on the D2-brane does from the scalar fluctuations of
D0-branes which are perpendicular to the fuzzy sphere.

\vspace{0.3cm}
\subsection{The effective action for the fuzzy plane}

\vspace{0.3cm}
In this subsection we construct the effective action for the fuzzy plane
by adding the fluctuations around this background.
As we have discussed in the section {\ref{sec:Comm}},
the matrices which represents the fuzzy plane is given by (\ref{eq:plamat}).
And there we defined $\rho_m = \frac{1}{2}(\rho_{m+1/2} + \rho_{m-1/2})$
and introduced the length $l_m$ so as to satisfy the relation
$2\pi \rho_m l_m = 2\pi a$.
This meant that the each area occupies the constant area $2\pi a$.
Here we interpret the area $2\pi a$ as the fuzziness of the D0-brane.

Now we also define the length $l_{m+1/2}$ so as to satisfy 
$2\pi \rho_{m+1/2} l_{m+1/2} = 2\pi a$.
The length $l_{m+1/2}$ is considered to represent the separation 
between the $m$th D0-brane and $(m+1)$th D0-brane, since the separation
is written as $\rho_{m+1} - \rho_m = 2a/(\sqrt{2a(m+1)} + \sqrt{2a(m-1)})$.
With these preparations we adopt the fluctuations around the fuzzy plane as
\begin{alignat}{3}
  X^1_{mn} &= \frac{1}{2} (\rho + l a_\phi)
  e^{il a_\rho} \Big|_{m+1/2} \delta_{m+1,n} 
  + \frac{1}{2} (\rho + l a_\phi)
  e^{- il a_\rho} \Big|_{m-1/2} \delta_{m,n+1} , \notag
  \\
  X^2_{mn} &= \frac{i}{2} (\rho + l a_\phi)
  e^{il a_\rho} \Big|_{m+1/2} \delta_{m+1,n} 
  - \frac{i}{2} (\rho + l a_\phi)
  e^{- il a_\rho} \Big|_{m-1/2} \delta_{m,n+1} , \label{eq:matpla}
  \\
  X^3_{mn} &= (z_0 + \hat{z}) \Big|_m \delta_{m,n} ,\qquad
  A_{t\,mn} = a_t \Big|_m \delta_{m,n} . \notag
\end{alignat}
Here the fluctuations $a_t|_m$, $a_\rho|_{m+1/2}$, $a_\phi|_{m+1/2}$ 
and $\hat{z}|_m$ depend on the time $t$.

Now let us evaluate the interior of the square root in the 
action (\ref{eq:D0}) step by step.
First the covariant derivatives are evaluated as
\begin{alignat}{3}
  (D_t X^1)_{mn} &= \frac{1}{2} \Big( \frac{a}{\rho} \dot{a}_\phi
  + ia f_{t\rho} + \mathcal{O}(\tfrac{l a_\phi}{\rho}) \Big)
  e^{il a_\rho} \Big|_{m+1/2} \delta_{m+1,n} \notag
  \\
  &\quad + \frac{1}{2} \Big( \frac{a}{\rho} \dot{a}_\phi
  - ia f_{t\rho} + \mathcal{O}(\tfrac{la_\phi}{\rho}) \Big)
  e^{-il a_\rho} \Big|_{m-1/2} \delta_{m,n+1}, \notag
  \\
  (D_t X^2)_{mn} &= \frac{i}{2} \Big( \frac{a}{\rho} \dot{a}_\phi
  + ia f_{t\rho} + \mathcal{O}(\tfrac{la_\phi}{\rho}) \Big)
  e^{il a_\rho} \Big|_{m+1/2} \delta_{m+1,n} 
  \\
  &\quad - \frac{i}{2} \Big( \frac{a}{\rho} \dot{a}_\phi
  - ia f_{t\rho} + \mathcal{O}(\tfrac{la_\phi}{\rho}) \Big)
  e^{-il a_\rho} \Big|_{m-1/2} \delta_{m,n+1}, \notag
  \\
  (D_t X^3)_{mn} &= \dot{\hat{z}} \Big|_m \delta_{m,n} , \notag
\end{alignat}
where $f_{t\rho}|_{m+1/2}$ is defined as 
$f_{t\rho}|_{m+1/2} = \dot{a}_\rho|_{m+1/2} - (a_t|_{m+1} - a_t|_m)/l_{m+1/2}$.
From these the second term in the square root is calculated as
\begin{alignat}{3}
  {(D_t X^i)^2}_{mn} &= \Big\{ \dot{z}^2 + a^2 \Big( \frac{1}{\rho^2} 
  \dot{a}_\phi^2 + f_{t\rho}^2 \Big) + \mathcal{O} (\tfrac{la_\phi}{\rho}) 
  \Big\} \Big|_m \delta_{m,n} ,
\end{alignat}
where $\frac{1}{\rho^2}\dot{a}_\phi^2|_m$ and $f_{t\rho}^2|_m$ are defined as
$\frac{1}{\rho^2}\dot{a}_\phi^2|_m = \frac{1}{2}(\frac{1}{\rho^2}
\dot{a}_\phi^2|_{m+1/2} + \frac{1}{\rho^2}\dot{a}_\phi^2|_{m-1/2})$ and 
$f_{t\rho}^2|_m = \frac{1}{2}(f_{t\rho}^2|_{m+1/2} + f_{t\rho}^2|_{m-1/2})$
respectively.

Second the commutators are estimated as
\begin{alignat}{3}
  \frac{1}{\lambda} [X^1,X^2]_{mn} &= \Big\{\! -\frac{ia^2}{\lambda \rho}
  \Big( \frac{\rho}{a} + a'_\phi \Big) + \mathcal{O}(\tfrac{la_\phi}{\rho}) 
  \Big\} \Big|_m \delta_{m,n}, \notag
  \\
  \frac{1}{\lambda} [X^2,X^3]_{mn} &= \frac{i}{2} \Big( \frac{a}{\lambda} 
  \hat{z}' + \mathcal{O} (\tfrac{la_\phi}{\rho}) \Big) 
  e^{il a_\rho} \Big|_{m+1/2} \delta_{m+1,n} \notag
  \\
  &\quad 
  + \frac{i}{2} \Big( \frac{a}{\lambda} \hat{z}' + \mathcal{O} 
  (\tfrac{la_\phi}{\rho}) \Big) e^{- il a_\rho} \Big|_{m-1/2} \delta_{m,n+1} ,
  \\
  \frac{1}{\lambda} [X^3,X^1]_{mn} &= - \frac{1}{2} \Big( \frac{a}{\lambda} 
  \hat{z}' + \mathcal{O} (\tfrac{la_\phi}{\rho}) \Big) 
  e^{il a_\rho} \Big|_{m+1/2} \delta_{m+1,n} \notag
  \\
  &\quad 
  + \frac{1}{2} \Big( \frac{a}{\lambda} \hat{z}' + \mathcal{O} 
  (\tfrac{la_\phi}{\rho}) \Big) e^{- il a_\rho} \Big|_{m-1/2} \delta_{m,n+1} , 
  \notag
\end{alignat}
where $a'_\phi|_m = (a_\phi|_{m+1/2} \!-\! a_\phi|_{m-1/2})/l_m$
and $\hat{z}'|_{m+1/2} = (\hat{z}|_{m+1} \!-\! \hat{z}|_m)/l_{m+1/2}$.
Then the third term in the square root becomes
\begin{alignat}{3}
  \Big(\frac{1}{2\lambda^2}[X^i,X^j]^2 \Big)_{mn} &=
  \Big\{\! - \frac{a^2}{\lambda^2} \hat{z}'^2
  - \frac{a^4}{\lambda^2 \rho^2} \Big( \frac{\rho}{a} + 
  a'_\phi \Big)^2 + \mathcal{O} (\tfrac{la_\phi}{\rho}) \Big\}
  \Big|_m \delta_{m,n} ,
\end{alignat}
where $\hat{z}'^2|_m$ is defined as
$\hat{z}'^2|_m = \frac{1}{2}(\hat{z}'^2|_{m+1/2}+\hat{z}'^2|_{m-1/2})$.

Third the matrix $\frac{\epsilon_{ijk}}{2\lambda} \{D_t X^i,[X^j,X^k]\}$ 
is calculated as
\begin{alignat}{3}
  \Big(\frac{\epsilon_{ijk}}{2\lambda} \big\{D_t X^i,[X^j,X^k] 
  \big\}\Big)_{mn} &= \Big[\! - \frac{ia^2}{\lambda \rho} 
  \Big\{ \dot{\hat{z}} \Big( \frac{\rho}{a} + a'_\phi \Big) 
  - \hat{z}' \dot{a}_\phi \Big\} + \mathcal{O} (\tfrac{la_\phi}{\rho}) 
  \Big] \Big|_m \delta_{m,n},
\end{alignat}
where $\frac{1}{\rho}\hat{z}'\dot{a}_\phi|_m$ is defined as
$\frac{1}{\rho}\hat{z}'\dot{a}_\phi|_m = \frac{1}{2}
(\frac{1}{\rho}\hat{z}'\dot{a}_\phi|_{m+1/2} + 
\frac{1}{\rho}\hat{z}'\dot{a}_\phi|_{m-1/2})$.
From this the fourth term in the square root is evaluated as
\begin{alignat}{3}
  \Big(\frac{\epsilon_{ijk}}{2\lambda} \big\{D_t X^i,[X^j,X^k] 
  \big\}\Big)^2_{mn} &= \Big[\! - \frac{a^4}{\lambda^2 \rho^2} 
  \Big\{ \dot{\hat{z}} \Big( \frac{\rho}{a} + a'_\phi \Big) 
  - \hat{z}' \dot{a}_\phi \Big\}^2 + \mathcal{O} (\tfrac{la_\phi}{\rho}) 
  \Big] \Big|_m \delta_{m,n}.
\end{alignat}

Now we are ready to evaluate the action (\ref{eq:D0}). 
From the calculations
so far, it is easy to see that the trace operation becomes just the 
summation since the the square root part is proportional to 
the unit matrix. Then we obtain the effective action of the form
\begin{alignat}{3}
  S_{\text{D}0} &= -T_0 \!\int\! dt \sum_{m=1}^\infty \frac{\rho l}{a}
  \Big[ 1 + \Big(\! -\dot{\hat{z}}^2 + \frac{a^2}{\lambda^2}\hat{z}'^2 \Big)
  + a^2 \Big\{\!- f_{t\rho}^2 
  + \frac{a^2}{\lambda^2 \rho^2} \Big( \frac{\rho}{a} + a'_\phi \Big)^2
  \!- \frac{1}{\rho^2} \dot{a}_\phi^2 \Big\} \notag
  \\
  &\qquad\qquad\qquad\qquad
  - \frac{a^4}{\lambda^2\rho^2} 
  \Big\{ \dot{\hat{z}} \Big( \frac{\rho}{a} + a'_\phi \Big) 
  - \hat{z}' \dot{a}_\phi \Big\}^2
  + \mathcal{O} (\tfrac{la_\phi}{\rho}) \Big]^{1/2} \Big|_m . \label{eq:D0pla}
\end{alignat}
This is the effective action for the fuzzy plane.

Let us consider the continuous limit of the above action.
When the fuzziness of the D0-brane $2\pi a$ is sufficiently small, 
it is possible to replace $\sum_m l_m$
with $\int d\rho$. Then the above action reaches to the form
\begin{alignat}{3}
  S_{\text{D}0} &= - \frac{T_0}{a} \!\int\! dt d\rho \, \rho
  \Big[ 1 + \Big(\! -\dot{\hat{z}}^2 + \frac{a^2}{\lambda^2}\hat{z}'^2 \Big)
  + a^2 \Big\{\!- f_{t\rho}^2 
  + \frac{a^2}{\lambda^2 \rho^2} \Big( \frac{\rho}{a} + a'_\phi \Big)^2
  \!- \frac{1}{\rho^2} \dot{a}_\phi^2 \Big\} \notag
  \\
  &\qquad\qquad\qquad\qquad
  - \frac{a^4}{\lambda^2\rho^2} 
  \Big\{ \dot{\hat{z}} \Big( \frac{\rho}{a} + a'_\phi \Big) 
  - \hat{z}' \dot{a}_\phi \Big\}^2
  + \mathcal{O} (\tfrac{a a_\phi}{\rho^2}) \Big]^{1/2} . \label{eq:D0plac}
\end{alignat}
This action precisely coincides with the 
action (\ref{eq:D2pla}) in the case of $a = b = \lambda$.
In other words, the matrices (\ref{eq:matpla}) correctly describe
the gauge and scalar fluctuations on the D2-brane under the 
condition of $a = b = \lambda$.

From the results obtained in this subsection, 
we conclude that the gauge fluctuations
on the D2-brane world-volume appear from the scalar fluctuations of
D0-branes which are parallel to the fuzzy plane and the scalar fluctuation
on the D2-brane does from the scalar fluctuations of
D0-branes which are perpendicular to the fuzzy plane.

\vspace{1cm}
\section{Conclusions and Discussions}

\vspace{0.3cm}
Through this paper we have investigated the correspondence between
the single D2-brane action, which is described by the abelian Born-Infeld 
action, and multiple D0-branes action,
which is done by the nonabelian Born-Infeld action.
And we found that these two actions precisely coincide with each other
if the condition $a = b = \lambda$ holds.

In the section \ref{sec:Comm}, we obtained the matrices which
describe the fuzzy cylinder, sphere, plane
and the general fuzzy surface with axial symmetry.
It should be emphasized that the explicit expressions of the matrices
(\ref{eq:cylmat}), (\ref{eq:surf}) ,(\ref{eq:sphmat}) and (\ref{eq:plamat})
make us possible to understand these fuzzy geometries pictorially
as Figs.\,\ref{fig:cyl}, \ref{fig:surf} and \ref{fig:sphpla}.

In the section \ref{sec:Effe}, the effective actions for the cylindrical,
spherical and planar D2-brane with magnetic flux on the world-volume 
are obtained by evaluating the abelian Born-Infeld action.
The results are given by (\ref{eq:D2cyl}), (\ref{eq:D2sph}) and
(\ref{eq:D2pla}), and the parameter $2\pi b$ introduced there represents
the area occupied per a unit of magnetic flux.

In the section \ref{sec:Const}, we have constructed the effective
actions for the fuzzy cylinder, sphere and plane
by evaluating the nonabelian Born-Infeld action for D0-branes. 
We started from the matrices
(\ref{eq:matcyl}), (\ref{eq:matsph}) and (\ref{eq:matpla}) and obtained
the results (\ref{eq:D0cyl}), (\ref{eq:D0sph}) and
(\ref{eq:D0pla}). The parameter $2\pi a$ introduced there 
represents the fuzziness of the D0-brane.
Our main result is that in the continuous limit
the effective actions reach to (\ref{eq:D0cylc}), (\ref{eq:D0sphc}) and
(\ref{eq:D0plac}), and those are precisely coincident with
the actions (\ref{eq:D2cyl}), (\ref{eq:D2sph}) and
(\ref{eq:D2pla}) respectively under the condition of $a = b = \lambda$.

Let us consider the physical meaning of the above conditions step by step.
First the condition $a = b$ is easy to understand if we note
that the charge of the D0-brane and the the magnetic flux on the D2-brane
are essentially the same quantum number.
It is interesting to mention that if we neglect the gauge
and scalar fluctuations, for example, the potential energy for 
the fuzzy cylinder precisely reaches to that for the cylindrical 
D2-brane in the continuous limit, only under the condition of $a = b$.

Next let us consider the meaning of $2\pi a = 2\pi \lambda$.
It is easy to see that the area $(2\pi\ell_s)^2$ is expressed as
\begin{alignat}{3}
  (2\pi\ell_s)^2 T_2 = T_0,
\end{alignat}
where $T_2$ is the tension of the D2-brane and $T_0$ is the mass of the
D0-brane. This means that the mass of the D2-brane with the 
area $(2\pi\ell_s)^2$ is equal to the mass of the D0-brane.
In other words, the D0-brane can transform into the D2-brane with the area
$(2\pi\ell_s)^2$ in view of the energy conservation.
Therefore it might be natural to conclude that
multiple D0-branes can describe the D2-brane if the condition
$2\pi a = 2\pi \lambda = (2\pi\ell_s)^2$ holds.

As a future work, it is interesting to apply our methods obtained in
this paper to more complicated cases. For example, it would be possible
to construct the nonabelian D2-branes action from multiple D0-branes
or the abelian D3-brane action from multiple D1-branes action.
It is also interesting to apply our methods to 
construct fundamental strings or a pair of brane anti-brane 
system\cite{Hya2,AHH}.

\vspace{1cm}
\section*{Acknowledgements}

\vspace{0.3cm}
I would like to thank Koji Hashimoto, Yasunari Kurita, Nobuyoshi Ohta 
and Norisuke Sakai for useful comments and members of the elementary particle
theory group in the Yukawa Institute for Theoretical Physics.
I thank the Yukawa Institute for Theoretical Physics 
at Kyoto University, where this work was developed during 
the YITP-W-02-19 on ``Extradimensions and Braneworld''.



\vspace{1cm}
\begin{thebibliography}{99}

\bibitem{Mye} R. C. Myers,
\textit{``Dielectric Branes''}, JHEP 9912 (1999) 022; hep-th/9910053.

\bibitem{BFSS} T. Banks, W. Fischler, S.H. Shenker and L. Susskind,
\textit{``M Theory as a Matrix Model: A Conjecture''},
Phys.Rev.D55:5112-5128,1997; hep-th/9610043.

\bibitem{VK} E. Keski-Vakkuri, P. Kraus
\textit{`` Born-Infeld Actions From Matrix Theory''}, \\
Nucl.Phys. \textbf{B518} (1998) 212-236; hep-th/9709122.

\bibitem{Hya} Y. Hyakutake, 
\textit{``Torus-like Dielectric D2-brane''}, \\
JHEP 0105 (2001) 013; hep-th/0103146.

\bibitem{BL} D. Bak, Ki-Myeong Lee, 
\textit{``Noncommutative Supersymmetric Tubes''}, \\
Phys.Lett.B509:168-174,2001; hep-th/0103148.

\bibitem{BSS} T. Banks, N. Seiberg, S. H. Shenker, 
\textit{``Branes from Matrices''}, \\
Nucl.Phys.B490:91-106,1997; hep-th/9612157.

\bibitem{Lei} R. G. Leigh, 
Mod.Phys.Lett. \textbf{A4} (1989) 2767.

\bibitem{Tse} A. A. Tseytlin, 
\textit{``Born-Infeld action, supersymmetry and string theory''}, \\
hep-th/9908105.

\bibitem{Tse2} A. A. Tseytlin,
\textit{``On Nonabelian Generalization of Born-Infeld Action 
in String Theory''}, Nucl.Phys.B501:41-52,1997; hep-th/9701125.

\bibitem{TR} W. Taylor, M. V. Raamsdonk, 
\textit{``Multiple Dp-branes in Weak Background Fields''}, \\
Nucl.Phys. \textbf{B573} (2000) 703-734; hep-th/9910052.

\bibitem{Hya2} Y. Hyakutake, 
\textit{``Expanded Strings in the Background of NS5-branes via a M2-brane,
a D2-brane and D0-branes''}, 
JHEP 0201 (2002) 021; hep-th/0112073.

\bibitem{AHH} H. Awata, S. Hirano and Y. Hyakutake, 
\textit{``Tachyon Condensation and Graviton Production in Matrix Theory''}, 
hep-th/99020158.

\bibitem{NO} S. Nojiri, S. D. Odintsov,
\textit{``Effective Potential for D-brane in Constant Electromagnetic Field''},
Int. J. Mod. Phys. A13 (1998) 2165; hep-th/9707142.

\bibitem{KNOS} T. Kadoyoshi, S. Nojiri, S. D. Odintsov and A. Sugamoto,
\textit{``Vacuun Polarization of Supersymmetric D-brane in the Constant 
Electromagnetic Field''},
Mod. Phys. Lett. A13 (1998) 1531; hep-th/9710010.

\bibitem{NO2} S. Nojiri, S. D. Odintsov,
\textit{``ON the Instability of Effective Potential for Nonabelian 
Toroidal D-brane''},
Phys. Lett. B419 (1998) 107; hep-th/9710137.

\bibitem{S} G. K. Savvidy,
\textit{``D0-branes with Nonzero Angular Momentum''},
hep-th/0108233.

\bibitem{SS} K. G. Savvidy, G. K. Savvidy,
\textit{``Stability of the Rotating Ellipsoidal D0-brane System''},
Phys. Lett. B501 (2001) 283; hep-th/0009029.

\bibitem{AST} T. Asakawa, S. Sugimoto and S. Terashima,
\textit{``Exact Description of D branes via Tachyon Condensation''},
hep-th/0212188.


\end{thebibliography}
\end{document}